\documentclass[12pt]{article}
\pdfoutput=1
\usepackage{jheppub}
\usepackage{amsmath}
\usepackage{MnSymbol}

\def\CI{{\cal I}}

\def\be{\begin{equation}}
\def\ee{\end{equation}}
\def\ba{\begin{aligned}}
\def\ea{\end{aligned}}
\def\ben{\begin{eqnarray}\displaystyle}
\def\een{\end{eqnarray}}

\def\nn{\nonumber}

\def\tq{{\tilde q}}

\newcommand{\bal}{\begin{equation}\begin{aligned}}
\newcommand{\eal}{\end{aligned}\end{equation}}

\preprint{
{\small{\textsf{DMUS-MP-13/07, CERN-PH-TH/2013-046}}}}

\title{3d \& 5d gauge theory partition functions as $q$-deformed CFT correlators}

\author[1]{Fabrizio Nieri}
\author[1]{Sara Pasquetti}
\affiliation[1]{Department of Mathematics, University of Surrey, Guildford, Surrey, GU2 7XH, UK}
\emailAdd{fb.nieri@gmail.com, sara.pasquetti@gmail.com, filipasse@gmail.com}

\author[2]{Filippo Passerini}
\affiliation[2]{PH-TH division, CERN, CH-1211 Geneva, Switzerland}

\abstract{
3d $\mathcal{N}=2$ partition functions on the squashed three-sphere $S^3_b$ and on the twisted product  $S^2\times S^1$ have been shown to factorize into  sums of squares of solid tori partition functions, the so-called holomorphic blocks.
The same set of holomorphic blocks realizes   $S^3_b$ and  $S^2\times S^1$ partition functions but
the two cases involve different inner products, the $S$-pairing and the $id$-pairing respectively.
We define a class of  $q$-deformed CFT correlators  where conformal blocks are controlled by  a deformation of Virasoro symmetry and are  paired by $S$-pairing and $id$-pairing respectively. Applying the bootstrap approach to a class of degenerate correlators  we are able to derive three-point functions.  We  show that  degenerate correlators can be  mapped to  3d partition functions while the crossing symmetry of CFT correlators corresponds to the flop symmetry of  3d  gauge theories.
We explore how non-degenerate $q$-deformed correlators are related to 5d partition functions.
We  argue  that $id$-pairing correlators are associated to the superconformal index
on $S^4\times S^1$ while  $S$-pairing three-point function factors capture the one-loop part 
of $S^5$ partition functions.
This is consistent with the interpretation of $S^2\times S^1$  and $S^3_b$   gauge theories as codimension two defect theories inside  $S^4\times S^1$  and $S^5$ respectively.
}

\begin{document}

\bibliographystyle{utphys}

\maketitle

\section{Introduction}

In recent years,  thanks to the application of the method of supersymmetric localization initiated by Pestun \cite{Pestun:2007rz},  several exact results for supersymmetric  theories formulated on compact manifolds have been obtained. In particular,  partition functions of $\mathcal{N}=2$ theories  on  the squashed three-sphere $S^3_b$ and  the superconformal index on   $S^2\times_q S^1$ (where $S^2$ is fibered  over $S^1$ with holonomy $\log q$) have been shown to localize to matrix integrals \cite{Kapustin:2009kz,Hama:2011ea,Kim:2009wb,Imamura:2011su,kw}.

In  \cite{Pasquetti:2011fj, hb} it was found that $S^3_b$ and  $S^2\times_q S^1$ partition functions can be factorized into   sums of squares of  partition functions on solid tori $D\times_q S^1$ (where $D\simeq \mathbb{R}^2$ is a cigar), named holomorphic blocks. For  any given $\mathcal{N}=2$ theory with $n$ isolated SUSY vacua, it is possible to compute the relevant  set of $n$ holomorphic blocks by means of an integral  formalism  developed in \cite{hb}.
Remarkably, the two partition functions are expressed in terms of the same set of blocks. However
$S^3_b$ and $S^2\times_q S^1$ partition functions are  obtained by fusing holomorphic blocks with different  inner products, which we call $S$-pairing and  $id$-pairing respectively.
The labeling of the pairings reflects the fact that  $S^3_b$ and  $S^2\times_q S^1$ 
are obtained by gluing solid tori through $S$ and $id$ element in $SL(2,\mathbb{Z})$ respectively.
Correspondently we will also refer to the partition functions  as $Z_S$ and $Z_{\rm id}$ respectively.
In  \cite{hb}  it was also shown that holomorphic blocks have an interesting behavior under a  certain class of mirror symmetry transformations. In order to guarantee invariance of the  partition function across mirror frames, holomorphic blocks
are constrained to undergo  subtle transformations across frames, which have been related to Stokes jumps.

In this work we focus on another symmetry of  $Z_S$ and $Z_{\rm id}$ partition functions, which we call flop symmetry  \cite{Witten:1993yc,Aspinwall:1993yb}, since it  exchanges phases of  the theory where the  Fayet-Iliopoulos (FI) parameter takes different signs. Also in this  case, 
the invariance of partition functions which  in the integral form is a rather trivial invariance of the integrand,
translates, in the block factorized form, into highly non-trivial transformation properties of blocks across phases.
One of the questions we try to answer in this paper is how much flop symmetry constrains  the form of partition functions.

This reasoning is reminiscent of the bootstrap approach to  2d CFT \cite{Belavin:1984vu}, where  correlation functions are constrained by crossing symmetry, that  follows from the  associativity of the operator algebra.   We review the bootstrap approach to   Liouville  CFT  in section \ref{boot}, where crossing symmetry  together with properties of degenerate representations of the Virasoro algebra constraints the structure of degenerate four-point functions and allows to determine the three-point function for generic primaries. This method  is commonly know as Teschner trick \cite{Teschner:1995yf}. We will then seek for a CFT realization of  our 3d gauge theory partition functions where
flop invariance is realized as crossing symmetry.

Recently a similar correspondence between gauge theory partition functions  on the two-sphere  and CFT correlators has been proposed in \cite{Doroud:2012xw}. Partition functions of  $\mathcal{N}=(2,2)$ theories on the two-sphere  have been computed with two different  localization schemes \cite{Doroud:2012xw, Benini:2012ui}. The first localization scheme reduces the path integral to an integral over the Coulomb branch of  classical action and one-loop fluctuations,  and is commonly referred to as the Coulomb branch localization.
In the other localization scheme, the so-called Higgs branch localization, the partition function takes the form 
of  a sum over Higgs vacua. Each term in the sum contains the classical action, one-loop fluctuations and a square of  vortex and anti-vortex excitations localized at the north and south poles of the sphere. Vortex partition functions   can be computed via equivariant localization, formulating the theory on  $\mathbb{R}^2$ with $\Omega$-deformation  with equivariant parameter $\epsilon$.
In \cite{Doroud:2012xw} it was shown  that the Higgs branch  $S^2$ partition function of  the SQED with $N_f$ fundamental chirals and $N_f$ anti-fundamental chirals is equivalent to  a four-point correlation function in $A_{N_f-1}$ Toda CFT,  where the insertions are a semi-degenerate state, two non-degenerate states and a completely degenerate state.  The authors provided a physical explanation of this relation using the AGT duality \cite{Alday:2009aq}(see also \cite{Wyllard:2009hg}), that relates  partition functions of 4d $\mathcal{N}=2$ $SU(N)$ gauge theories to correlation functions of $A_{N-1}$ Toda CFT.  Indeed, coupling  the  2d SQED  gauge theory to the 4d superconformal QCD, gives rise to a certain defect surface operator for the 4d theory, described in the AGT set up by a  degenerate insertion \cite{Alday:2009fs}. It follows that in the limit where the 2d SQED theory decouples from the 4d superconformal  QCD, the partition for the 2d theory is described by the above mentioned correlation function in Toda CFT. 

Although to date, only the Coulomb branch  localization scheme for 3d partition function is known, as we will review in section \ref{3dpf},  both $Z_S$ and $Z_{\rm id}$ can be recast in a form very similar to the $S^2$ Higgs branch localization, that is a sum over SUSY vacua of   classical and  one-loop terms and a square of vortex partition functions which are now paired respectively with $S$-pairing and  $id$-pairing. In the 3d case the vortex partition function is the natural $q$-deformation of the two-dimensional case, and can be computed  by K-theory equivariant localization  formulating the theory on $\mathbb{R}^2\times S^1$ with equivariant parameter $q=e^{\beta \epsilon}$, with $\beta$ the $S^1$ length.

Motivated by the strong similarities between 2d and 3d partition functions,  we then  study a class of  CFT correlators  where conformal blocks are controlled by  a $q$-deformation of the $\mathcal{W}_{N}$ algebra \cite{Feigin:1995sf, Awata:1995zk}. For simplicity we will restrict to the  $N=2$ case (\emph{i.e.}  the CFT related to 3d SQED gauge theory with four flavors), but most of our results can be trivially extended to the $N>2$ case.  We will therefore focus on the so-called $\mathcal{V}ir_{q,t}$ algebra, a deformation of the Virasoro algebra. Indeed this algebra,  introduced and developed in \cite{Shiraishi:1995rp},  was also studied in connection with the $5$d extension of the AGT conjecture \cite{Awata:2009ur,Awata:2010yy,Schiappa:2009cc,Mironov:2011dk,Yanagida:2010vz},
since it appears as the natural   deformation corresponding to the 5d-lift of the Nekrasov instanton partition function. 
In particular,   it has been checked that  K-theory  instanton partition functions are reproduced by $q$-deformed conformal blocks.

Our  goal is however to study correlation functions rather than simply chiral conformal blocks and here comes the first novelty: how do we pair $q$-deformed conformal blocks? In fact we should ask which property of the correlation function
we want our pairing to be compatible with. We will  require modular invariance (crossing symmetry) of $q$-correlators and try to find compatible ways of pairing $q$-deformed conformal blocks.
The results in 3d gauge theory  suggest to use $S$-pairing and $id$-pairing (and in principle pairing with any  $S(2,\mathbb{Z})$ elements), since they are  compatible with flop symmetry which we think of as the gauge theory analogous of crossing symmetry.

In sections \ref{qboot}, \ref{idsec}, \ref{ssec} we  develop a  $q$-deformed version of the bootstrap approach to Liouville
which allows us to determine  three-point functions for correlators involving $S$-pairing and $id$-pairing respectively.
Using these three-point functions we compute four-point correlation functions with three non-degenerate and one degenerate primary and show that they can be mapped to  $Z_S$ (partition function on squashed sphere) and $Z_{\rm id}$ (superconformal index) of 3d SQED with $N_f=2$, in analogy with the 2d case.

We then turn to correlators of four non-degenerate primaries focusing on the three-point function contribution.
Here comes an interesting surprise. If we use three-point functions that we derived considering the $id$-pairing,  we obtain the one-loop part of the $SU(2)$ $N_f=4$ theory on $S^4\times S^1$ recently computed in \cite{Kim:2012gu,Terashima:2012ra, Iqbal:2012xm}, 
while if we use three-point functions for the $S$-pairing we obtain the one-loop part of the
$SU(2)$ $N_f=4$ theory on squashed $S^5$ recently derived in \cite{Imamura:2012bm, Lockhart:2012vp, Kim:2012qf} extending results in \cite{Kallen:2012cs, Hosomichi:2012ek, Kallen:2012va, Kim:2012ava}.

Perhaps the first result is not so surprising since it is the obvious  5d lift of the  AGT
correspondence,  and the $S^2\times_q S^1$ theory  that  corresponds to a degenerate $id$-correlator, can be interpreted as the dimensional uplift  of a  surface operator.
The relation of  $S$-correlators with $S^5$ partition function is more intriguing.
 As described in \cite{Lockhart:2012vp} and reviewed in section \ref{s5}, it is useful to  view the squashed  $S^5$ as a $T^3$ fibration over a triangle. Over each edge the fiber  degenerates to a $T^2$ fibration leading to three  squashed three-spheres  inside $S^5$. It is then natural,  in analogy with the AGT scenario, 
to associate the defect theories defined on these three maximal three-spheres  to degenerate CFT correlators.

To summarize, our results suggest that, not only the 5d Nekrasov instanton function is related to $q$-deformed $W$-algebra, but also  full partition functions  of 5d gauge theories on $S^4\times S^1$ and $S^5$ can be mapped to interacting theories with $q$-deformed $W$ symmetry. Like in the AGT correspondence, the 3d $S^2\times_q S^1$ and $S^3_b$ partition functions,  captured by degenerate correlators, describe the partition function of certain codimension two defect theories.

\section{3d partition functions}
\label{3dpf}

In this section we will study the partition function of the $\mathcal{N}=2$  SQED  with $N_f$ fundamental chirals and $N_f$ anti-fundamental chirals on $S^2\times_q S^1$ (\emph{i.e.} the superconformal index) and $S^3_b$.
We will begin by reviewing the block-factorization property \cite{Pasquetti:2011fj,hb} (see also \cite{Hwang:2012jh}).
 We will then show the non-trivial constraints imposed by flop symmetry on the holomorphic blocks.

 \subsection{The superconformal index factorization}
 In this section we study the $\mathcal{N}=2$ SQED, the $U(1)$ theory with $N_f$ fundamental chirals and $N_f$ anti-fundamental chirals on the twisted product  $S^2\times_q S^1$. The path integral on this manifold defines a superconformal  index  for the theory and was shown to reduce to a finite dimensional integral in \cite{Kim:2009wb, Imamura:2011su}  using supersymmetric localization. A generalization of the index that allows background fields with non-trivial magnetic flux was introduced in \cite{kw} and further developed in \cite{DGG2}\footnote{For a derivation of the superconformal index  that involves the index theorem see \cite{Drukker:2012sr}.}.
 We turn on fugacities (together with their magnetic flux through $S^2$) for all the flavor symmetries and the topological $U(1)$ symmetry, that corresponds to the Fayet-Iliopoulos (FI) paramater. Our fugacities are:
\ben\nonumber
&&(\phi_i, r_i), \qquad i=1,\cdots N_f,\qquad {\rm flavor}\quad U(1)^{N_f}\,,\\ \nonumber
&& (\xi_i, l_i), \qquad i=1,\cdots N_f, \qquad {\rm (anti)-flavor}\quad U(1)^{N_f}\,, \\ \nonumber
&& (\omega, n), \qquad {\rm topological}\quad U(1)\,,\\
&& (t, s), \qquad {\rm gauged}\quad U(1)\,.
\een
The 1-loop contribution of a single chiral multiplet of R-charge 0 is \cite{DGG2}
\be
\label{chi}
\chi(\zeta,m)=(q^{1/2} \zeta^{-1})^{-m/2} \CI_\Delta(\zeta,m)
\ee
where $\CI_\Delta(\zeta,m)$ is the index of a free chiral with $k=-1/2$ Chern Simons (CS) units \cite{DGG2}.
\footnote{In the 3d-3d correspondence  this is the  theory associated to the ideal tetrahedron  \cite{DGG1, DGG2}.}
 It is given by 
\be
\CI_\Delta(\zeta,m)=\prod_{k=0}^\infty\frac{(1-q^{l+1} \zeta^{-1}q^{-m/2}   )}{(1-q^{l} \zeta
q^{-m/2}   )}=\prod_{k=0}^\infty\frac{(1-q^{l+1} x^{-1}   )}{(1-q^{l} \tilde x^{-1}
 )}=\Big|\Big| (q x^{-1};q)_\infty \Big|\Big|^2_{\rm id}  \, ,
\ee
where we defined $x=\zeta q^{m/2}$ and $\tilde x= \zeta^{-1} q^{m/2}$.
We take $\zeta$ to be a phase, $m \in \mathbb{Z}$ and  $q$ real,
so that $\tilde x= \bar x$.\footnote{The bar denotes complex conjugation.} We also defined the identity pairing
\be
\Big|\Big| f(x;q) \Big|\Big|^2_{\rm id}:= f(x;q)f(\tilde x;\tilde q)\,,
\ee 
with $\tilde q=q^{-1}$.
The coupling of the topological $U(1)$ current to background fields produces the following classical contribution
\be
t^n \omega^s\,.
\ee
The index  is computed using a Coulomb branch localization scheme, where the path integral reduces to the integration over saddle points labeled by a holonomy  and a quantized magnetic flux for the dynamical gauge vector. It is given by 
\be
\label{cind}
 Z_{\rm id}= \sum_{s\in \mathbb{Z}} \int \frac{d t}{2\pi i t} t^n \omega^s \prod_{j=1}^{N_f}\chi(t\phi_j,s+r_j)
\prod_{k=1}^{N_f}\chi(t^{-1}\xi_k^{-1},-s-l_k)\, .
\ee

To evaluate the integral (\ref{cind}) we take the contribution of poles inside the unit circle,
coming from the (denominators) of the fundamental hypers, that are located at
  \ben
t= \phi_i^{-1} q^{(s+r_i)/2} q^{-k} , && \quad k\geq min(0,s+r_i), \qquad i=1,\cdots N_f
\, . \een
We refer the reader to the Appendix \ref{iid} for details of the computation, here we just give the final result.
We first introduce some notation
\ben
 x_i= \phi_i q^{r_i/2}, \qquad \tilde x_i= \phi_i^{-1} q^{r_i/2}\, ,\qquad
 y_i= \xi_i q^{l_i/2}, \qquad \tilde y_i= \xi_i^{-1} q^{l_i/2}\, , \quad  z=\omega q^{n/2}
 \een
 and
 \be \prod_{j,k}^{N_f}x_j^{-1} y_k=r=e^R;\quad  u=(-q^\frac{1}{2})^{N_f}r^\frac{1}{2}z^{-1}\,.
\ee 
We also introduce the following theta function
\be
 \theta( x ;q )\equiv( - q^{1/2} x;q )_\infty (- q^{1/2} x^{-1};q )_\infty\,,
\ee
which satisfies
\be
|| \theta((- q^{1/2})^c x^a;q )||^2_{id}=
(- q^{1/2} )^{-(a\cdot m) c} \zeta^{-(a\cdot m) a}\,,
\ee
where $x_i=\zeta_i q^{m_i/2}$ and $a_i$ is a vector of $N$ integers and $c$ is an integer. 

The index can be written as
\ben
\label{hiin}
Z_{\rm id}=\sum_{i=1}^{N_f} G^{(i)}_{cl}  G^{(i)}_{1loop}  \Big|\Big| Z_V^{i}  \Big|\Big|^2_{\rm id} \,,
\een
where the various factors are given by:
 \be
G^{(i)}_{cl}=\omega^{-r_i}(  \phi_i^{-1})^n =\Big|\Big| \frac{\theta(z x_i ;q)}{\theta(z ;q) \theta(x_i ;q)} \Big|\Big|^2_{\rm id}\,,
\ee

\ben\nn
 G^{(i)}_{1loop}&=&\prod_{j=1}^{N_f}\prod_{k=1}^{N_f}
\Big|\Big| (q x_i x_j^{-1};q)_\infty  (q  y_j x_i^{-1};q)_\infty  \Big|\Big|^2_{\rm id} (q^{1/2} \phi_i \phi_j^{-1})^{(r_i-r_j)/2}
(q^{1/2} \xi_k \phi_i^{-1})^{(l_k-r_i)/2}
 =\\
&&=
\prod_{j=1}^{N_f}\chi(\phi_j \phi_i^{-1},r_j-r_i) \prod_{k=1}^{N_f} \chi( \phi_i \xi_k^{-1},r_i-l_k) \,,
\een
and
\be
Z_V^{(i)}=\sum_p \prod_{j,k=1}^{N_f} \frac{ ( x_i y_k^{-1};q)_{p}}{(q x_i x_j^{-1};q)_{p}}u^{p}=\phantom{|}_{N_f}\Phi_{N_f-1}(x_i y_1^{-1},\ldots,x_i y_{N_f}^{-1};q x_i x_1^{-1},\hat\ldots,q x_i x_{N_f}^{-1};u) \,,
\ee
where $\phantom{|}_{n+1}\Phi_{n}$ is a basic hypergeometric function defined in (\ref{qhyper}) and the hat means that the $i$-th entry is omitted. Notice that $G^{(i)}_{cl}$ and  $G^{(i)}_{1loop}$ are  exactly of the same form of the  classical term and 1-loop term in equation (\ref{cind}), which is derived via Coulomb branch localization. In particular, $G^{(i)}_{cl}$ and  $G^{(i)}_{1loop}$ are equivalent to the Coulomb branch terms with  coulomb branch parameters fixed as $t=\phi_i^{-1}$ and $s=-r_i$.   Hence (\ref{hiin}) is the  form one would expect to obtain using an alternative localization scheme corresponding to  the Higgs branch localization. The equivalence between the factorized partition functions and the Higgs branch localization has been discussed for 2d gauge theories in \cite{Doroud:2012xw, Benini:2012ui}.

We can also write the partition function in terms of holomorphic blocks  as in \cite{hb}.
We have:\footnote{Up to a prefactor
 independent on $i$.}
\be
G^{(i)}_{cl}G^{(i)}_{1loop}\Big| \Big| 
  \mathcal{Z}_V^{(i)}\Big| \Big|_{\rm id}^2  =
 \Big|\Big|   \frac{\theta(x_i u;q)}{\theta(u;q)\theta(x_i;q)} \prod_{j,k}^{N_f} \frac{(q x_j x_i^{-1};q)_\infty}{( y_k x_i^{-1};q)_\infty }  \Big|\Big|^2_{\rm id}
 \Big| \Big| \mathcal{Z}^{(i)}_{V} \Big| \Big|_{\rm id}^2:= \Big| \Big| \mathcal{B}_i\Big| \Big|_{\rm id}^2
\ee
and
\be
\label{inbl}
Z_{\rm id}=  \sum_{i=1}^{N_f}  \Big| \Big| \mathcal{B}_i\Big| \Big|_{\rm id}^2\, .
\ee

\subsection{Ellipsoid partition function factorization}
We will now turn to the ellipsoid SQED partition function. We turn on   masses $m_i$ for the $N_f$ chirals,  masses $\tilde m_i$ for the $N_f$ anti-chiral and  an FI parameter $\xi$. The ellipsoid partition function was computed in \cite{Hama:2011ea}, generalizing previous results for the round $S^3$ \cite{Kapustin:2009kz, Jafferis:2010un, Hama:2010av}\footnote{A derivation of the ellipsoid partition function by the index theorem is described in \cite{Drukker:2012sr}.}. It reads
\be
\label{elli}
Z_S=\int dx ~e^{2\pi i x \xi}~ \prod_{j,k}^{N_f}~\frac{s_b(x+m_j+ i Q/2)}{s_b(x+\tilde m_k- i Q/2)}\,,
\ee
where $Q=b+1/b$ and $s_b(x)$ is the double-sine function described in Appendix \ref{3sap}.  To evaluate the integral we close the contour in the upper half plane and take the contributions of poles
located at $x=-m_i + i m b + i n/b$,  see \cite{Pasquetti:2011fj} for details.
As before we first introduce some notation: 
\ben
\label{etil}\nn
x_i&=&e^{2\pi b m_i}, \quad y_i=e^{2\pi b \tilde m_i}, \quad z=e^{2\pi b \xi}, \quad q=e^{2\pi i b^2}\,,\\
\tilde x_i&=&e^{2\pi  m_i/b}, \quad \tilde y_i=e^{2\pi  \tilde m_i/b}, \quad \tilde z=e^{2\pi \xi/b}, \quad \tilde q=e^{2\pi i/ b^2} \,,
\een
and
\be \prod_{j,k}^{N_f}x_jy_k^{-1}=r,\quad  u=(-q^\frac{1}{2})^{N_f}r^\frac{1}{2}z^{-1}\,.
\ee
We also define the $S$-pairing \cite{hb}
\be
\label{sp}
\Big| \Big| f(x;q)\Big| \Big|_{S}^2=f(x;q)f(\tilde x;\tilde q)\,,
\ee
where all the variables are defined as in equation (\ref{etil}).
In particular we have
\be 
\Big| \Big| \theta\big( (-q^{1/2})^{c} x^{a};q\big)\Big| \Big|_{S}^2 = {\rm C}^{-2} \exp\Big[-i \pi\Big(a \frac{\log x}{2 i\pi b}+c \frac{Q}{2}\Big)^2\Big]\,,
\ee
where ${\rm C}=e^{-\frac{i\pi}{12}(b^2+\frac{1}{b^2})}$. The result reads
\be\label{Zellipsoid}
Z_S= \sum_i^{N_f} G^{(i)}_{cl}  G^{(i)}_{1-loop} \Big|\Big|\mathcal{Z}^{(i)}_V\Big|\Big|^2_S\,,
\ee
where the various terms are given by 
\be
\label{cl3p}
G^{(i)}_{cl}=e^{-2\pi i \xi m_i}=
\Big|\Big| \frac{\theta(z x_i^{-1} ;q)}{\theta(z ;q) \theta(x_i^{-1} ;q)} \Big|\Big|^2_{S}\,, \qquad G^{(i)}_{1-loop}=\prod_{j,k}^{N_f} \frac{s_b(m_j-m_i+i Q/2)}{s_b(\tilde m_k-m_i-i Q/2)}\, ,
\ee
\ben
\mathcal{Z}^{(i)}_V=\sum_n\prod_{j,k}^{N_f} \frac{( y_k x_i^{-1};q)_n}{(q x_j x_i^{-1};q)_n}  u^n=\phantom{|}_{N_f}\Phi_{N_f-1}(x_i^{-1}y_1,\ldots,x_i^{-1} y_{N_f};q x_i^{-1} x_1,\hat\ldots,q x_i ^{-1}x_{N_f};u)\, .\qquad
\een
$\phantom{|}_{N_f}\Phi_{N_f-1}$ is a basic hypergeometric function defined in (\ref{qhyper}).  As in the case of the superconformal index,  $G^{(i)}_{cl}$ and $G^{(i)}_{1-loop}$ are equivalent to the classical and 1-loop contribution that appear in the Coulomb branch localization formula (\ref{elli}). In particular, they are equivalent to the Coulomb branch factors with the Coulomb branch parameter fixed as $x=-m_i$. The expression (\ref{Zellipsoid}) is therefore expected to follow from a Higgs branch localization scheme, similar to the 2d case \cite{Doroud:2012xw, Benini:2012ui}.

We can further write the partition function in terms of holomorphic blocks  as in \cite{hb}:\footnote{Up to a prefactor
 independent on $i$.}
\be
G^{(i)}_{cl}G^{(i)}_{1-loop}\Big| \Big| 
  \mathcal{Z}_V^{(i)}\Big| \Big|_{S}^2  =
 \Big|\Big|   \frac{\theta(x_i u;q)}{\theta(u;q)\theta(x_i;q)} \prod_{j,k}^{N_f} \frac{(q x_j x_i^{-1};q)_\infty}{( y_k x_i^{-1};q)_\infty }  \Big|\Big|^2_S
 \Big| \Big| \mathcal{Z}^{(i)}_{V} \Big| \Big|_{S}^2:= \Big| \Big| \mathcal{B}_i\Big| \Big|_{S}^2\,,
\ee
with
\be
Z_S=  \sum_{i=1}^{N_f}  \Big| \Big| \mathcal{B}_i\Big| \Big|_{S}^2\, ,
\ee
with exactly the same blocks we obtained for the index in (\ref{inbl}) (with $x_i \to x_i^{-1}$ and $y_i \to y_i^{-1}$).

\subsection{Flop invariance}
We now come to the main point of this section:
index and ellipsoid partition functions are invariant under flop symmetry which
swaps phase $I$ and phase $II$ of the theory corresponding to positive and negative FI parameter.
At the level of the integral form of partition functions, 
flop symmetry boils down to a very simple invariance of the integrand.  Namely, 
the superconformal index in equation (\ref{cind}) is invariant under the transformations $\omega  \leftrightarrow  \omega^{-1}$,  $n \leftrightarrow   -n$, $\phi_j \leftrightarrow  \xi_j^{-1} $, $r_j \leftrightarrow  - l_j .$ 
 Similarly,  the ellipsoid partition function in equation  (\ref{elli})  is  invariant under $m_i \leftrightarrow -\tilde m_k$, $\xi \leftrightarrow -\xi$, as it follows using that  $s_b(x)s_b(-x)=1$.

Flop symmetry  translates into highly non-trivial relations between Higgs branch quantities (\ref{hiin}), (\ref{Zellipsoid}) in phase $I$, $G^{(i),I}_{cl},  G^{(i),I}_{1loop},  \mathcal{Z}_v^{(i),I}, $  and the corresponding phase $II$ quantities. Namely  we have:
\ben
\label{flopin}
Z_{\rm id}(x_j,y_k, z)=Z_{\rm id}^I\!\!&=&\!\!\sum_i^{N_f}G^{(i),I}_{cl}  G^{(i),I}_{1loop}  \Big|\Big| \mathcal{Z}_V^{(i),I}  \Big|\Big|^2_{\rm id}= \nn \\&&\!\!\!\!\!\!\!= \sum_i^{N_f}
G^{(i),II}_{cl}  G^{(i),II}_{1loop}  \Big|\Big| \mathcal{Z}_V^{(i),II}  \Big|\Big|^2_{\rm id}=Z_{\rm id}(y_k^{-1},x_j^{-1}, z^{-1})=Z_{\rm id}^{II}\,,
\een
and
\ben
\label{flopel}
Z_{S}(x_j,y_k, z) =Z_{S}^I\!\!&=&\!\!\sum_i^{N_f}G^{(i),I}_{cl}  G^{(i),I}_{1loop}  \Big|\Big| \mathcal{Z}_V^{(i),I}  \Big|\Big|^2_{S}= \nn \\&&\!\!\!\!\!\!\!= \sum_i^{N_f}
G^{(i),II}_{cl}  G^{(i),II}_{1loop}  \Big|\Big| \mathcal{Z}_V^{(i),II}  \Big|\Big|^2_{S}=Z_{S}(y_k^{-1},x_j^{-1}, z^{-1})=Z_{S}^{II}\,.
\een
We have checked these relations but we will not write down details in this section as in the next section
we will perform a very similar computation. We only mention that to check these equations one can for example analytically continue the $q$-series in $ \mathcal{Z}_V^{(i),I}$ from phase $I$ to phase $II$ using a generalisation of  equation (\ref{cont}) and then identify the coefficients of
$  \Big|\Big| \mathcal{Z}_V^{(i),II}  \Big|\Big|^2_{S,id}$ on the two sides of the equality. 
An important point to be aware of is that because of  the $S$-pairing and  the $id$-pairing
involve  $q$-series with $|q|<1$ and $|\tilde q|>1$ we need to use appropriate  analytic continuations
in the two regimes, as it was pointed out in \cite{hb}.

The symmetry we have just described resembles the way associativity of the operator algebra in a 2d CFT constrains the form of correlation functions. It is then very natural to  wonder whether it is possible to map the  3d partition functions  to CFT correlators, so that flop symmetry is guaranteed by  crossing symmetry of the CFT.  A similar correspondence holds for 2d gauge theories, where the partition function results to be equivalent to a degenerate correlator in Toda CFT \cite{Doroud:2012xw}.

\section{$q-$deformed CFT   correlation functions}

In this section we study a class of CFT correlation functions with   degenerate insertions, where conformal blocks are fixed by the $\mathcal{V}ir_{q,t}$  symmetry (\emph{i.e.} a $q$-deformation of Virasoro) and are paired  requiring the modular invariance of the correlation function. It turns out that both the $id$-pairing and the $S$-pairing,  defined in the previous section,
are compatible with modular invariance and three-point correlation functions can be computed  via the bootstrap approach.

We start reviewing the bootstrap approach \cite{Belavin:1984vu} to Liouville theory, showing how the associativity of the operator algebra and properties of degenerate representations of Virasoro algebra constraint the structure of degenerate four-point functions and determine the  three-point function for generic primaries \cite{Belavin:1984vu, Dotsenko:1984ad, Dotsenko:1984nm, Teschner:1995yf, Pakman:2006hm}. 
 
We then consider a $q$-deformed version of the bootstrap procedure,  deriving  the structure of the $q$-deformed degenerate four-point functions and the  $q$-deformed three-point function for each pairing.

\subsection{The bootstrap approach to Liouville theory}\label{boot}
Liouville theory is a non-rational CFT whose primary fields $V_\alpha(z, \tilde{z})$  are  labeled by the  momentum $\alpha$.\footnote{Many properties of Liouville CFT are described for instance in \cite{Nakayama:2004vk, Teschner:2001rv}. } The conformal dimension of the primary $V_\alpha(z, \tilde{z})$ is given by $\Delta_\alpha=\alpha(Q_0-\alpha)$ and the Virasoro central charge is  $c_V=1+6Q_0^2$, where $Q_0=b_0+\frac{1}{b_0}$ and  $b_0\in \mathbb{R}^+$.  The theory admits a Lagrangian description in terms of a two dimensional scalar field and in this language, the parameter $b_0$ is a field theory coupling constant. In the following we will not use any Lagrangian formulation, and review how the Liouville three-point function can be derived simply exploiting the bootstrap approach \cite{Teschner:1995yf}.  

\begin{figure}[!ht]
\leavevmode
\begin{center}
\includegraphics[height=2.7cm]{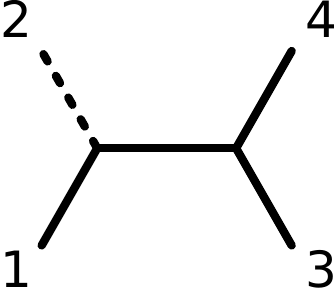}
\end{center}
\caption{The $s$-channel decomposition of the correlator given in formula (\ref{scorre}). The dashed line is associated to the degenerate state $\alpha_2=-\frac{b_0}{2}$.}
\label{schannel}
\end{figure}

The non-degenerate representations of the Virasoro algebra  correspond to primaries with momentum $\alpha=\frac{Q_0}{2}+i p_\alpha$, where $p_\alpha\in  \mathbb{R}^+$. Degenerate representations are labeled by two positive integers $n$ and $m$, and a degenerate representation with a null state at level $nm$ is associated to a primary field with momentum $\alpha^{(m,n)}=\frac{Q_0}{2}-\frac{m}{2b_0}-\frac{nb_0}{2}$. We are interested in the four-point correlation function 
\ben\label{lfour}
\langle V_{\alpha_4}(z_4,\tilde z_4) V_{\alpha_3}(z_3,\tilde z_3) V_{\alpha_2}(z_2, \tilde z_2) V_{\alpha_1}(z_1,\tilde z_1) \rangle\, ,
\een
 where  $z$ and $\tilde z$ are complex coordinates that can be considered as independent for the moment.  Due to the projective Ward identities, the correlation function assumes the following form  
\ben
&&\langle V_{\alpha_4}(z_4,\tilde z_4) V_{\alpha_3}(z_3, \tilde z_3) V_{\alpha_2}(z_2, \tilde z_2) V_{\alpha_1}(z_1, \tilde z_1) \rangle \\ \nn
&&\quad=
  (z_{24})^{-2\Delta_2} (z_{14})^{\Delta_2+\Delta_3-\Delta_1-\Delta_4} (z_{34})^{\Delta_1+\Delta_2-\Delta_3-\Delta_4} (z_{13})^{\Delta_4-\Delta_1-\Delta_2-\Delta_3}\\ \nn
   &&\qquad\times (\tilde{z}_{24})^{-2\Delta_2} (\tilde{z}_{14})^{\Delta_2+\Delta_3-\Delta_1-\Delta_4} (\tilde{z}_{34})^{\Delta_1+\Delta_2-\Delta_3-\Delta_4} (\tilde{z}_{13})^{\Delta_4-\Delta_1-\Delta_2-\Delta_3}\,H (z,\tilde{z})\,,
 \een
where $z=\frac{z_{12}z_{34}}{z_{13}z_{24}}$,  $z_{ij}=z_i-z_j$ and similar definitions hold for tilded variables. Considering the primary $V_{\alpha_2}(z_2, \tilde z_2)$ to be  degenerate  with a null state at level 2, that is with degenerate Liouville momentum $\alpha_2=\alpha^{(1,2)}=-\frac{b_0}{2}$,\footnote{Also the representation with $\alpha^{(1,2)}=-\frac{1}{2b_0}$ has a null state at level 2. This is in agreement with the fact that Liouville theory is invariant under the exchange $b_0\leftrightarrow \frac{1}{b_0}$.}  the correlation function satisfies a second order differential equation \cite{Belavin:1984vu}. Using projective invariance to fix the position of three operators as $z_4=\infty$, $z_3=1$, $z_1=0$ and similar for tilded variables,  the differential equation reads 
\be\label{bpzeq}
 R(\Delta_1,\Delta_2,\Delta_3,\Delta_4; z) H(z,\tilde{z})=0\,,\qquad
  R(\Delta_1,\Delta_2,\Delta_3,\Delta_4;\tilde z) H(z,\tilde{z})=0\,,
\ee
where the differential operator is given by 
\ben\nn
&& R(\Delta_1,\Delta_2,\Delta_3,\Delta_4; z) =\\ &&\qquad \frac{3}{2(2\Delta_2+1)}\frac{\partial^2}{\partial z^2} + \left( \frac{1}{z-1} + \frac{1}{z} \right) \frac{\partial}{\partial z}
-\frac{\Delta_3}{(z-1)^2} - \frac{\Delta_1}{z^2} + \frac{\Delta_1 + \Delta_2 + \Delta_3 -\Delta_4 }{z(z-1)}\,.\qquad
\een
We remark that $z,\tilde z$ describe the position of the degenerate operator $V_{\alpha_2}$. It is convenient to introduce the function $G(z,\tilde z )$ related to $H(z,\tilde{z})$ by\footnote{A similar parameterization for the conformal blocks has been used in the context of  $A_{N-1}$ Toda CFT in \cite{Fateev:2005gs, Fateev:2007ab}.}
\ben\label{cb}
H(z,\tilde{z})=z^{b_0\alpha_1}(1-z)^{b_0\alpha_3}\,\tilde z^{\,b_0\alpha_1}(1-\tilde z)^{b_0\alpha_3}G(z,\tilde z )   
\een
so that the (\ref{bpzeq}) can be written as  
\ben\label{hyeq}
D(a,b;c; z) G(z,\tilde{z})=0\,,\qquad
 D(a,b;c;\tilde z) G(z,\tilde{z})=0\,,
\een
where $D(a,b;c; z)$ is the hypergeometric differential operator. It is given by 
\ben\label{hyop}
D(a,b;c; z) =z(1-z)\frac{\partial^2}{\partial z ^2}+[c-(a+b+1)z])\frac{\partial}{\partial z }-ab
\een
and the parameters $a,b,c$ are related to Liouville momenta as
\ben\label{parmom}
a&=&-1+b_0(\alpha_1+\alpha_3+\alpha_4-3b_0/2)\,,\nn\\
b&=&b_0(\alpha_1+\alpha_3-\alpha_4-b_0/2)\,,\nn\\
c&=&2\alpha_1b_0-b_0^2\,.
\een
The function $G(z,\tilde{z})$ is therefore a bilinear combination of solutions of hypergeometric equations, in the variables $z$ and $\tilde z$. It is known that such solutions are singular at the points $0,1,\infty$  that in the CFT language correspond to the locations of  the three non-degenerate fields. We start analyzing the hypergeometric differential equation  in a punctured neighborhood of the point  $0$. In this region, two linearly independent solutions are given by  
\ben \label{zerosol}
I_1^{(s)}(z)=\phantom{|}_2F_1(a,b;c;z),\quad\qquad I_2^{(s)}(z)=z^{1-c}\phantom{|}_2F_1(1+a-c,1+b-c;2-c;z)\,,
\een
and analog solutions are found for the equation in the $\tilde z$ variable. The  general solution of the equations (\ref{hyeq}) is given by  $G(z,\tilde{z})=\sum_{i,j=1}^2I_i^{(s)}(\tilde z) K_{ij}^{(s)}I_j^{(s)}(z)$  for a generic constant matrix $K_{ij}^{(s)}$. The vector of solutions $(I_1^{(s)},I_2^{(s)})$ (\ref{zerosol}) is not single valued when the singularity at $0$  is encircled, but it transforms by the element of the monodromy group associated to the singularity in $0$, {\em i.e.} $I_i^{(s)}\rightarrow \sum_{j=1}^2Y^{(s)}_{(0)ij}I_j^{(s)}$.  In order to construct a single valued correlation function,  we now  use the fact that  Liouville theory is defined on  2d Euclidean space, implying that the two holomorphic coordinates are related by complex conjugation, i.e. $\tilde z=\bar z$. Given that the monodromy matrix $Y^{(s)}_{(0)ij}$ is diagonal\footnote{A vector of solutions has diagonal monodromy matrix around a singularity included in the domain where the solutions are defined.} and unitary, a single valued correlation function is obtained considering a diagonal matrix  $K^{(s)}_{ij}$.  Reminding  that $0$ is the position of the primary  $V_{\alpha_1}$ and that $z$ is the position of the degenerate field $V_{-\frac{b_0}{2}}$,  it follows that the conformal blocks defined nearby  $0$ correspond to the  $s$-channel decomposition of the correlation function, see Figure \ref{schannel}.
We  therefore write the $s$-channel degenerate correlation function  as 
\ben\label{scorre}
\langle V_{\alpha_4}(\infty) V_{\alpha_3}(1) V_{\alpha_2}(z) V_{\alpha_1}(0) \rangle&=&|z|^{2b_0\alpha_1}|1-z|^{2b_0\alpha_3}\sum_{i,j=1}^2I_i^{(s)}(\bar z) K_{ij}^{(s)}I_j^{(s)}(z)\\ \nn
&=&|z|^{2b_0\alpha_1}|1-z|^{2b_0\alpha_3}\sum_{i=1}^{2}C(\alpha_4,\alpha_3,\beta_i^{(s)})\,C_i(\alpha_1) \Big| \Big| I_i^{(s)}(z) \Big| \Big| ^2
\een
where the elements of the diagonal matrix $K^{(s)}_{ij}$ have been written in terms of  the three point correlation functions  $C(Q_0-\alpha,\beta,\gamma)\propto \langle \alpha | V_{\beta}(z,\bar z)|\gamma \rangle$. More explicitly 
\ben\label{ks}
K_{11}^{(s)}=C(\alpha_4,\alpha_3,\beta_1^{(s)})\,C_1(\alpha_1),\quad K_{22}^{(s)}=C(\alpha_4,\alpha_3,\beta_2^{(s)})\,C_2(\alpha_1),\quad K_{12}^{(s)}=K_{21}^{(s)}=0\,,\qquad
\een
where   $\beta_i^{(s)}$ are the momenta of the internal states in the two fusion channels. They are given by 
\be
\beta_1^{(s)}=\alpha_1-\frac{b_0}{2}\, , \quad \beta_2^{(s)}=\alpha_1+\frac{b_0}{2}
\ee
 as expected by the fusion rules of the primary $V_{\alpha_1}$ with the degenerate state $V_{-\frac{b_0}{2}}$. We have defined also  
\ben
C_1(\alpha)=C\left(Q_0-(\alpha-\frac{b_0}{2}),-\frac{b_0}{2},\alpha\right)\, ,\quad C_2(\alpha)=C\left(Q_0-(\alpha+\frac{b_0}{2}),-\frac{b_0}{2},\alpha\right)\,.
\een
We have introduced the pairing\footnote{Note that this is different from the standard complex modulus squared since the hypergeometric solutions depend on complex parameters $a,b,c$ and therefore $( f(a,b,c,z))^*=f(\bar a,\bar b,\bar c, \bar z)$. This is because the hypergeometric solutions equal  the conformal blocks only up to prefactors,  as it follows form (\ref{cb}).}
 \be
 \label{unpa}
 \Big| \Big| f(a,b,c,z) \Big| \Big| ^2=f(a,b,c,z)f(a,b,c,\bar z)\,.
 \ee
In the following section, chiral CFT sectors will be coupled using the pairing we encountered in the factorized 3d gauge theories.
 
\begin{figure}[!ht]
\leavevmode
\begin{center}
\includegraphics[height=2.7cm]{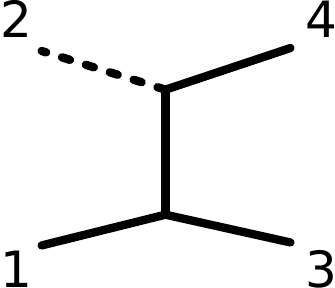}
\end{center}
\caption{The $u$-channel decomposition of the correlator given in formula (\ref{ucorre}). The dashed line is associated to the degenerate state $\alpha_2=-\frac{b_0}{2}$.}
\label{uchannel}
\end{figure}

We now consider a representation of the conformal blocks in the neighborhood of $z=\infty$, {\em i.e.} we  construct a solution of the differential equations (\ref{hyeq}) that is well defined for large $z$.  A set of independent solutions of the hypergeometric equation in the neighborhood of $z=\infty$ is given by 
\ben \label{inftysol}\nn
I_1^{(u)}(z)=z^{-a}\phantom{|}_2F_1(a,1+a-c;1+a-b;z^{-1}),\quad I_2^{(u)}(z)=z^{-b}\phantom{|}_2F_1(b,1+b-c;1+b-a;z^{-1})\,.\\
\een
Given that $z=\infty$ is the location of the $V_{\alpha_4}$ operator,  it follows that these conformal blocks are associated to the $u$-channel decomposition of the partition function, see Figure \ref{uchannel}. 
Also in this case, the solutions to the hypergeometric equations are not invariant under monodromy. In particular, when the singularity at infinity is encircled, the vector of solutions $(I_1^{(u)},I_2^{(u)})$ (\ref{inftysol}) transforms by a representation  of the element of the monodromy group associated to the singularity at infinity, {\em i.e.} $I_i^{(u)}\rightarrow \sum_{j=1}^2Y^{(u)}_{(\infty)ij}I_j^{(u)}$. Given that $Y^{(u)}_{(\infty)ij}$ is diagonal,  it results that a singled valued correlation function is given by 
\ben\label{ucorre}
\langle V_{\alpha_4}(\infty) V_{\alpha_3}(1) V_{\alpha_2}(z) V_{\alpha_1}(0) \rangle &=&|z|^{2b_0\alpha_1}|1-z|^{2b_0\alpha_3}\sum_{i,j=1}^2I_i^{(u)}(\bar z) K_{ij}^{(u)}I_j^{(u)}(z)\\ \nn
&=&|z|^{2b_0\alpha_1}|1-z|^{2b_0\alpha_3}\sum_{i=1}^{2}C(\alpha_1,\alpha_3,\beta_i^{(u)})\,C_i(\alpha_4)\Big| \Big|I_i^{(u)}(z)\Big| \Big|^2
\een
where the internal channel  states now are
\be
\beta_1^{(u)}=\alpha_4-\frac{b_0}{2}\, , \quad \beta_2^{(u)}=\alpha_4+\frac{b_0}{2}\,.
\ee
 The explicit expression for the matrix $K_{ij}^{(u)}$ is given by 
\ben\label{ku}
K_{11}^{(u)}=C(\alpha_1,\alpha_3,\beta_1^{(u)})\,C_1(\alpha_4),\quad K_{22}^{(u)}=C(\alpha_1,\alpha_3,\beta_2^{(u)})\,C_2(\alpha_4),\quad K_{12}^{(u)}=K_{21}^{(u)}=0.\qquad
\een

We have constructed the correlation function in the $s$-channel (\ref{scorre}), that is defined for small values of $z$ and the correlation function in the $u$-channel (\ref{ucorre}),  that is defined for large values of $z$. We have simply considered  set of solutions defined in the relevant domain and coupled the holomorphic and anti-holomorphic part in such a way to have a single valued correlation function. 

It is also possible to extend a set of solutions outside their domain of definition using analytical continuation. For instance, using the analytical continuation of the hypergeometric function  
 \ben
\phantom{|}_2F_1(a,b;c;z)=&& \frac{\Gamma(c) \Gamma(b-a)}{\Gamma(b) \Gamma(c-a)} (-z)^{-a} \phantom{|}_2F_1(a,1-c+a;1-b+a; z^{-1})  \nn \\
&&+ \frac{\Gamma(c) \Gamma(a-b)}{\Gamma(a) \Gamma(c-b)} (-z)^{-b} \phantom{|}_2F_1(b,1-c+b;1-a+b; z^{-1})\,,
\een 
it follows that the solutions $I_i^{(s)}$ (\ref{zerosol}), when analytically continued outside the domain $|z|<1$, are linearly related to the solutions $I_i^{(u)}$ (\ref{inftysol}) as 
\ben
I_i^{(s)}=\sum_{j=1}^2M_{ij}I_j^{(u)}\,,
\een 
where the elements of the matrix $M_{ij}$ are given by 
\ben\nn\label{con}
M_{11}=\frac{\Gamma(c) \Gamma(b-a)}{\Gamma(b) \Gamma(c-a)}\,,\qquad  M_{12}= \frac{\Gamma(c) \Gamma(a-b)}{\Gamma(a) \Gamma(c-b)}\,,\\
M_{21}=\frac{\Gamma(2-c) \Gamma(b-a)}{\Gamma(1+b-c) \Gamma(1-a)}\,,\qquad  M_{22}= \frac{\Gamma(2-c) \Gamma(a-b)}{\Gamma(1+a-c) \Gamma(1-b)}\,.
\een
For consistency,  the analytical continuation of the correlation function in the $s$-channel (\ref{scorre}) is required  to be equivalent to  the correlation function in the $u$-channel (\ref{ucorre}) \cite{Belavin:1984vu, Dotsenko:1984ad, Dotsenko:1984nm,Teschner:1995yf},  {\em i.e.}
 \be\label{masterl}
K_{11}^{(s)} \Big| \Big| I^{(s)}_1  \Big| \Big|_{}^2+
K_{22}^{(s)}
\Big| \Big| I^{(s)}_2  \Big| \Big|_{}^2=
K_{11}^{(u)}
\Big| \Big| I^{(u)}_1  \Big| \Big|_{}^2+
K_{22}^{(u)}
\Big| \Big| I^{(u)}_2  \Big| \Big|_{}^2
\ee
that implies the following  matrix equation to be satisfied
\ben\label{master}
\sum_{k,l=1}^2K_{kl}^{(s)}M_{ki}M_{lj}=K^{(u)}_{ij}\,.
\een
It results that equation   (\ref{master})  determines the Liouville three-point function \cite{Teschner:1995yf, Pakman:2006hm}, only up to some trivial prefactors that has to be fixed using the Lagrangian formalism. The off-diagonal terms produce the following equation
\ben\label{k22}
\frac{K_{22}^{(s)}}{K_{11}^{(s)}}=-\frac{M_{11}M_{12}}{M_{21}M_{22}}\,,
\een
which using (\ref{ks}) and (\ref{con}) gives
\ben\label{tt}
\frac{C(\alpha_4,\alpha_3,\beta_2^{(s)})}{C(\alpha_4,\alpha_3,\beta_1^{(s)})}=\frac{C_1(\alpha_1)}{C_{2}(\alpha_1)}\frac{\gamma(c)\gamma(1-b)\gamma(1-c+a)}{\gamma(2-c)\gamma(c-b)\gamma(a)}\,,
\een
where $\gamma(x)=\frac{\Gamma(x)}{\Gamma(1-x)}$ and the parameters $a,b,c$ are related to the Liouville momenta as in  (\ref{parmom}).  

As in \cite{Pakman:2006hm}, we now consider a diagonal  term of the equation (\ref{master}), we take $(i=2,j=2)$. Plugging in (\ref{k22}) we obtain
\ben\label{diag}
\frac{K_{22}^{(u)}}{K_{22}^{(s)}}=(M_{22})^2-\frac{M_{21}M_{12}M_{22}}{M_{11}}=\frac{M_{22}}{M_{11}}(\det M)
\een
and using (\ref{ku}) and (\ref{con}) we have 
\ben\label{pak}
\frac{C(\alpha_1,\alpha_3,\beta_2^{(u)})\,C_2(\alpha_4)}{C(\alpha_4,\alpha_3,\beta_2^{(s)})\,C_2(\alpha_1)}=\gamma(b)\gamma(a-b)\gamma(c-a)\gamma(2-c)\,.
\een
This last equation (\ref{pak}), together with a normalization choice for primaries can be combined with  equation (\ref{tt}) to produce a difference equation that determine the three-point function $C(\alpha,\beta,\gamma)$ up to a prefactor that can be computed using the Lagrangian realization of Liouville \cite{Teschner:1995yf, Pakman:2006hm}. 

In view of what we will do in the next section, here we focus on the equations (\ref{tt}) and (\ref{pak}) and see what it can be learned form them without any other assumption. Defining $2\alpha_T = \alpha_1 + \alpha_2 + \alpha_3$, it is possible to show that the ansatz
\ben\label{dozz}
C(\alpha_1,\alpha_2,\alpha_3) =
\frac{1}{\Upsilon( 2\alpha_T - Q_0)} \prod_{r=1}^3 \frac{\Upsilon(2 \alpha_r)}{\Upsilon(2\alpha_T-2\alpha_r)} \,,
\een
where the function $\Upsilon(X)$ is defined in (\ref{nupsi}), solves both the equations  (\ref{tt}) and (\ref{pak}).  Up to a prefactor, the three-point function (\ref{dozz}) is in perfect agreement with the DOZZ expression \cite{Dorn:1994xn, Zamolodchikov:1995aa} and has been determined completely using the bootstrap approach.  We  notice that the expression (\ref{dozz}), although it is not the full DOZZ three-point function, it is the only part that is reproduced by a gauge theory one-loop computation in the AGT correspondence.  

In view of this result, in the next section we will focus on a $q$-analog of the equations (\ref{tt}) and (\ref{pak}) and use them to compute three point functions that will be reproduced by gauge theory computations.

\subsection{$q-$deforming the  bootstrap}\label{qboot}
In the previous section, the  Liouville degenerate four point function and the generic three point function were derived without any use of the Lagrangian. This was possible thanks to  the constraints imposed by degenerate representations of the Virasoro algebra, that is the symmetry of Liouville CFT. 

In this section we consider  a non-rational CFT whose primary fields  are associated to representations of a $q$-deformation of the Virasoro algebra introduced in \cite{Shiraishi:1995rp} .   The  $q$-Virasoro algebra  $\mathcal{V}ir_{q,t}$  has two complex parameters $q,t$ and it is useful also to consider their ratio $p=\frac{q}{t}$. There is an infinite set of  generators $T_n$ with $n\in \mathbb{Z}$ that satisfy the following commutation relation 
\ben
[T_n \, , \, T_m]=-\sum_{l=1}^{+\infty}f_l\left(T_{n-l}T_{m+l}-T_{m-l}T_{n+l}\right)
-\frac{(1-q)(1-t^{-1})}{1-p}(p^{n}-p^{-n})\delta_{m+n,0}\,,
\een
with  $f_{l}$  associated to the expansion of the function $f(z)$, \emph{i.e.}
\ben
f(z)=\sum_{l=0}^{+\infty}f_l z^l
=\exp \left[\sum_{l=1}^{+\infty}\frac{1}{n}\frac{(1-q^n)(1-t^{-n})}{1+p^n}
 z^n \right]\,.
\een
The algebra  $\mathcal{V}ir_{q,t}$  is invariant under the following transformations \cite{Shiraishi:1995rp, Awata:1996fq}
\ben\label{dsym}
(q,t)\rightarrow (q^{-1},t^{-1})\,,\qquad\qquad (q,t)\rightarrow (t,q)\, .
\een
Like for the Virasoro algebra,  representations of $\mathcal{V}ir_{q,t}$ can be constructed using Verma modules \cite{Shiraishi:1995rp}. The highest weight state $|\lambda\rangle$ satisfies 
\ben
T_0 |\lambda\rangle=\lambda |\lambda\rangle,\qquad T_n |\lambda\rangle=0\quad\text{for}\quad n>0,
\een
and the Verma module  $\mathcal{M}(\lambda)$ is constructed acting on the highest weight state $|\lambda\rangle$ with the operators $T_{-n}$  with  $n>0$.  Singular states in the Verma module can be detected using the Kac determinant. In particular, it is possible to show that there is a level two singular vector for the following values of the parameter $\lambda$
\ben\label{ddeg}
\lambda_1=p^{1/2}q^{1/2}+p^{-1/2}q^{-1/2},\qquad\qquad \lambda_2=p^{1/2}t^{-1/2}+p^{-1/2}t^{1/2}.
\een
We point out that the states  $\lambda_1$ and $\lambda_2$ are mapped into each other by the exchange $(q,t)\rightarrow (t,q)$ and they are left invariant by $(q,t)\rightarrow (q^{-1},t^{-1})$. The algebra $\mathcal{V}ir_{q,t}$ can be related to other known algebras when the parameters  $p,q$ assume certain specific values.\footnote{See \cite{Awata:1996fq} for an overview.} In particular, considering
\ben\label{lvir}
t=q^{-b_0^2}\qquad\text{and}\qquad q\rightarrow 1\,,
\een 
$\mathcal{V}ir_{q,t}$ reduces to the Virasoro algebra  with central charge $c_V=1+6Q_0^2$ where $Q_0=b_0+\frac{1}{b_0}$, that is the symmetry algebra of Liouville theory with coupling constant $b_0$. We note that the $(q,t)\rightarrow (t,q)$ symmetry of  $\mathcal{V}ir_{q,t}$  reduces to the $b_0\leftrightarrow \frac{1}{b_0}$ Virasoro/Liouville symmetry.  It is therefore natural to identify the states $\lambda_1$,$\lambda_2$ (\ref{ddeg}) as the $q$-deformation of the degenerate states $\alpha^{(1,2)}=-\frac{b_0}{2}$ and $\alpha^{(2,1)}=-\frac{1}{2b_0}$.

We now consider a non-rational CFT whose symmetry algebra is given by   tensor products of $\mathcal{V}ir_{q,t}$ and can be thought as a $q$-deformation of Liouville CFT.  We are interested in four-point correlation functions where three of the insertions are non-degenerate primaries, \emph{i.e.}  associated to non-degenerate representations of $\mathcal{V}ir_{q,t}$, and one of the insertion is associated to one of the degenerate representations in (\ref{ddeg}). Like for the Virasoro case, the degenerate state imposes constraints on the correlator. Using a bosonic representation of the algebra is possible to argue that the degenerate  chiral correlator satisfies a  $q$-hypergeometric difference equation \cite{Awata:1996xt, Awata:2010yy}. The same conclusion can be achieved studying a  $q$-deformation of the $\beta$ ensemble \cite{Schiappa:2009cc}. 

The correlation function we are interested in is a $q$-deformation of the four-point function described in details in the previous section, \emph{i.e.}  (\ref{lfour}). Like for the undeformed case, we assume  primaries to be labeled by  continuous parameters $\alpha_i$ and we take the primary $V_{\alpha_2}$ to be associated to a degenerate representation with a null state at level 2. Therefore we take the four-point $q$-deformed correlator to be
\ben
\langle V_{\alpha_4}(\infty) V_{\alpha_3}(r) V_{\alpha_2}(z) V_{\alpha_1}(0) \rangle&\sim& G(z,\tilde z),
\een
where  we omit a conformal prefactor and the function $G(z,\tilde z)$  satisfies a difference equation.  In particular  
\ben\label{qdif}
 D(A,B;C; q; z)G(z, z)=0\,,\qquad
 D(\tilde A,\tilde B;\tilde C;\tilde q; \tilde z)G(z,\tilde z)=0\,,
\een
where $D(A,B;C;q;z)$ is the $q$-hypergeometric  operator that is given by \cite{grahman}
\be 
 D(A,B;C;q;z)= h_2\,\frac{\partial_q^2}{\partial_q z^2}+h_1\frac{\partial_q}{\partial_q z}+h_0
\ee
where  $\frac{\partial_q}{\partial_q z}$ is the $q$-derivative that acts on a function $f(z)$ as
\ben
\frac{\partial_q}{\partial_q z}\, f(z)=\frac{f(qz)-f(z) }{z(q-1)}
\een
and $h_2,h_1,h_0$  are defined by
\ben\nn
h_2&=&z(C-AB q z),\\ \nn
h_1&=&\frac{1-C}{1-q}+\frac{(1-A)(1-B)-(1-AB q) }{(1-q)}z,\\
h_0&=&-\frac{(1-A)(1-B)}{(1-q)^2}.
\een
Non-degenerate primaries are inserted at  singular points $0,r,\infty$, where $r=\frac{q^{-1}C}{ AB}$, of the $q$-hypergeometric  operator.
The parameters $A,B,C$ are related to $\alpha_1,\alpha_3,\alpha_4$, however, as we will discuss in the next section,  the precise dictionary depends on  the pairing that it is used to glue the different chiral sectors.

We now investigate the constraints imposed on the four point function by the difference equation (\ref{qdif}); for the moment, we can consider the tilded variables as independent from the untilded ones.  As in the undeformed case, equation (\ref{qdif})  implies that  $G(z,\tilde z)$ is expressed as a linear combination of solutions of the $q$-hypergeometric difference equation.   A  basis of two  linearly independent  solutions   (with $|q|<1$)  in the neighborhood of $z=0$ is given by 
\be
I^{(s)}_1=  F^{(s)}_1(z),\qquad
I^{(s)}_2=T^{(s)}_2 F^{(s)}_2(z)\,,
\ee
where 
\ben \label{qzerosol}
F^{(s)}_1(z)=\phantom{|}_2\Phi_1(A,B;C;z),\quad\qquad F^{(s)}_2(z)=\phantom{|}_2\Phi_1(q A C^{-1},q B C^{-1};q^2 C^{-1};z)\nn\, .\\
\een
 $\phantom{|}_2\Phi_1(A,B;C;z)$ is the hypergeometric $q$-series defined in (\ref{qhyper21})
and
\be
T^{(s)}_{2}:=T_{qC^{-1}}( z^{-1} r^{1/2} q)
 \, ,\ee
and we introduced the twist function:
\be
T_A(u)=\frac{\theta( A u;q)}{\theta( A   ;q)\theta(u;q)}\,,
\ee 
which satisfies $T_A(q^n u)=(A)^{-n} T_A(u)$.
\footnote{It is easy to verify  that $D(A,B;C;q;z) T^{(s)}_2\sim  D(qA C^{-1}, qB C^{-1}; q^2C^{-1};q;z)$.} 
Notice that since
\ben
\lim_{q\to 1}\phantom{|}_2\Phi_1(q^{a},q^{b};q^{c};q,z)&=&\phantom{|}_2F_1(a,b;c;z)
\een 
and
\be
\lim_{q\to 1}T_A(u)=u^{-a}\,,
\ee
in the undeformed limit we recover the basis of $s$-channel solutions (\ref{zerosol}).

In analogy with the undeformed case we then construct the  $s$-channel degenerate correlation function as the following inner product of solutions defined in the neighborhood of $z=0$, \emph{i.e.} 
\ben\label{scorreq}
\langle V_{\alpha_4}(\infty) V_{\alpha_3}(r) V_{\alpha_2}(z) V_{\alpha_1}(0) \rangle&\sim&\sum_{i,j=1}^2 \tilde I_i^{(s)}(\tilde z;\tilde q) K_{ij}^{(s)}I_j^{(s)}(z; q)\\ \nn
&&=\sum_{i=1}^{2}K_{ii}^{(s)} \Big| \Big| I_i^{(s)}(z;q) \Big| \Big|_{*} ^2\,,
\een
where the elements of the diagonal matrix $K^{(s)}_{ij}$ can be interpreted as products of three point functions and 
we  defined the generic pairing   of $q$-deformed chiral sectors 
\be
\Big| \Big| f(A,B,C;z;q) \Big| \Big|_{*} ^2=f(A,B,C;z;q)f(\tilde A,\tilde B,\tilde C;\tilde z;\tilde q)\,.
\ee
In the next sections we will consider two different pairing,  inspired by the 3d block factorization, which allow to realise crossing symmetry invariant correlation functions.

The $u$-channel correlation function is obtained considering solutions of the deformed hypergeometric equation in the neighborhood of $z=\infty$. A basis of independent solutions in this domain is given by 
\ben
I^{(u)}_1=T^{(u)}_1 F^{(u)}_1(z^{-1} r q^2)
\, , \qquad 
I^{(u)}_2= 
T^{(u)}_2 F^{(u)}_2(z^{-1} r q^2)\,,
\een
where
\ben \label{qinftysol}\nn
F^{(u)}_1(z^{-1} r q^2)&=&\phantom{|}_2\Phi_1(A, q A C^{-1};q A B^{-1};q^2r z^{-1})\, ,\\
F^{(u)}_2(z^{-1} r q^2)&=&\phantom{|}_2\Phi_1(B,q B C^{-1};q B A^{-1};q^2 r z^{-1})\,,\nn\\
\een
and the $u$-channel twist functions are given by
\be
T^{(u)}_{1}:=T_{A^{-1}}(z^{-1} r^{1/2} q)\, , \quad
T^{(u)}_{2}:=T_{B^{-1}}(z^{-1} r^{1/2} q)\,.
\ee
Also in this case, in the $q\to 1$ limit we recover the undeformed $u$-channel basis of solutions  (\ref{inftysol}).
 The correlation function in the $u$-channel is therefore written as 
\ben\label{ucorreq}
\langle V_{\alpha_4}(\infty) V_{\alpha_3}(r) V_{\alpha_2}(z) V_{\alpha_1}(0) \rangle&\sim&\sum_{i,j=1}^2 \tilde I_i^{(u)}(\tilde z;\tilde q) K_{ij}^{(u)}I_j^{(u)}(z; q)\\ \nn
&&=\sum_{i=1}^{2}K_{ii}^{(u)} \Big| \Big| I_i^{(u)}(z;q) \Big| \Big|_{*} ^2\,.
\een

To construct a modular invariant object, as in the undeformed case, we demand crossing symmetry, which requires
 \be
 \label{qcs}
K_{11}^{(s)} \Big| \Big| I^{(s)}_1  \Big| \Big|_{*}^2+
K_{22}^{(s)}
\Big| \Big| I^{(s)}_2  \Big| \Big|_{*}^2=
K_{11}^{(u)}
\Big| \Big| I^{(u)}_1  \Big| \Big|_{*}^2+
K_{22}^{(u)}
\Big| \Big| I^{(u)}_2  \Big| \Big|_{*}^2\,,
\ee
where functions outside their domain of definition are defined via analytical continuation. In the following we will analytically continue the solutions $I^{(s)}_i(q;z)$ outside the domain $|z|<1$ and use equation (\ref{qcs}) to obtain  non-trivial equations for the matrices $K_{ij}^{(s)}$ and   $K_{ij}^{(u)}$. These equations are used in the next section to determine the $q$-deformed three point functions. 

We use the analytic continuation of the basic hypergeometric  series (\ref{cont}) to find\footnote{Notice that here, unlike in the undeformed case,  matrices $B_{ij}$ transform hypergeometric series rather than solutions of the hypergeometric equation.}
\be
\label{x1}
\Big| \Big| F^{(s)}_1(z)\Big| \Big|_{*}^2=
\left( B_{11} F^{(u)}_1(z^{-1} r q^2)+
B_{12} F^{{(u)}}_2(z^{-1} r q^2)\right)
\left(\tilde B_{11} \tilde F^{(u)}_1(z^{-1} r q^2)+\tilde B_{12}
\tilde F^{(u)}_2(z^{-1} r q^2)\right)\, ,
\ee
and
\be
\label{x2}
\Big| \Big| F^{(s)}_2(z)\Big| \Big|_{*}^2=
\left( B_{21} F^{(u)}_1(z^{-1} r q^2)+
B_{22} F^{{(u)}}_2(z^{-1} r q^2)\right)
\left(\tilde B_{21} \tilde F^{(u)}_1(z^{-1} r q^2)+\tilde B_{22}
\tilde F^{(u)}_2(z^{-1} r q^2)\right)\, ,
\ee
with
\ben\nonumber
B_{11}&=&\frac{(B;q)_\infty(CA^{-1};q)_\infty}{(C;q)_\infty(BA^{-1};q)_\infty}\frac{\theta_{11}(A^{-1}z^{-1};q)}{\theta_{11}(z^{-1};q)}\\ \nonumber
B_{12}&=&\frac{(A;q)_\infty(CB^{-1};q)_\infty}{(C;q)_\infty(AB^{-1};q)_\infty}\frac{\theta_{11}(B^{-1}z^{-1};q)}{\theta_{11}(z^{-1};q)}\\
\nonumber
B_{21}&=&\frac{(qBC^{-1};q)_\infty (qA^{-1};q)_\infty}{(q^2C^{-1};q)_\infty(BA^{-1};q)_\infty}\frac{\theta_{11}(q^{-1}CA^{-1}z^{-1};q)}{\theta_{11}(z^{-1});q}\\
\label{mb}
B_{22}&=&\frac{(qAC^{-1};q)_\infty(qB^{-1};q)_\infty}{(q^2C^{-1};q)_\infty(AB^{-1};q)_\infty}\frac{\theta_{11}(q^{-1}CB^{-1}z^{-1};q)}{\theta_{11}(z^{-1};q)}\, ,
\een
where $\theta_{11}(x;q)=\theta(-q^{1/2}x;q)$.
We need two different  analytic continuations of basic hypergeometric series, depending on whether  we are inside or outside the unit circle. Assuming $|q|<1$ we will have $|\tilde q|>1$, so we need another set of matrices:
\ben
\nn\tilde{B}_{11}&=&\frac{(\tq \tilde{C}^{-1};\tq)_\infty (\tq \tilde{A}\tilde{B}^{-1};\tq)_\infty}{(\tq \tilde{B}^{-1};\tq)_\infty (\tq\tilde{A}\tilde{C}^{-1};\tq)_\infty}
\frac{\theta_{11}(\tilde{C}\tilde{B}^{-1}\tilde{A}^{-1}\tilde{z}^{-1};\tq)}{\theta_{11}(\tilde{C}\tilde{B}^{-1}\tilde{z}^{-1};\tq)}\\\nn
\tilde{B}_{12}&=&\frac{(\tq \tilde{C}^{-1};\tq)_\infty(\tq \tilde{B}\tilde{A}^{-1};\tq)_\infty}{(\tq \tilde{A}^{-1};\tq)_\infty(\tq\tilde{B}\tilde{C}^{-1};\tq)_\infty}
\frac{\theta_{11}(\tilde{C}\tilde{A}^{-1}\tilde{B}^{-1}\tilde{z}^{-1};\tq)}{\theta_{11}(\tilde{C}\tilde{A}^{-1}\tilde{z}^{-1};\tq)}\\\nn
\tilde{B}_{21}&=&\frac{(\tq^{-1}\tilde{C};\tq)_\infty(\tq \tilde{A}\tilde{B}^{-1};\tq)_\infty}{(\tilde{C}\tilde{B}^{-1};\tq)_\infty(\tilde{A};\tq)_\infty}
\frac{\theta_{11}(\tilde{C}\tilde{B}^{-1}\tilde{A}^{-1}\tilde{z}^{-1};\tq)}{\theta_{11}(\tq\tilde{B}^{-1}\tilde{z}^{-1};\tq)}\\
\tilde{B}_{22}&=&\frac{(\tq^{-1}\tilde{C};\tq)_\infty(\tq \tilde{B}\tilde{A}^{-1};\tq)_\infty}{(\tilde{C}\tilde{A}^{-1};\tq)_\infty(\tilde{B};\tq)_\infty}
\frac{\theta_{11}(\tilde{C}\tilde{A}^{-1}\tilde{B}^{-1}\tilde{z}^{-1};\tq)}{\theta_{11}(\tq\tilde{A}^{-1}\tilde{z}^{-1};\tq)}.
\een
Inserting equations (\ref{x1}), (\ref{x2}) in equation  (\ref{qcs}) we derive two equations for three point function:
\begin{enumerate}
\item[i.] Imposing the vanishing of the cross-terms we get
\be
\label{qt1}
K^{(s)}_{11} B_{11}\tilde B_{12}+ K^{(s)}_{22}   \Big| \Big| T^{(s)}_2\Big| \Big|_{*}^2  B_{21}\tilde B_{22}=0\, .
\ee
Inserting equations (\ref{mb}) we find the following equation for the ratio of three point functions:

\be
\label{feq}
\frac{K^{(s)}_{22} }{K^{(s)}_{11} }=\Big| \Big| \frac{(A;q)_\infty (B;q)_\infty(q^2 C^{-1};q)_\infty  }{(C;q)_\infty (q AC^{-1};q)_\infty (q BC^{-1};q)_\infty  } \frac{\theta_{11}(q^{-1}A^{-1} C;q)\theta_{11}(A^{-1}z^{-1};q)}{\theta_{11}(A^{-1};q)\theta_{11}(q^{-1}A^{-1} C z^{-1};q)} \Big| \Big|_{*}^2 \Big| \Big|
\frac{1}{T^{(s)}_{2}}\Big| \Big|_{*}^2\,.
\ee

\item[ii.] Matching diagonal terms proportional to $F^{(u)}_2$ we get
\label{m2}
\be
K^{(s)}_{11}  B_{12} \tilde B_{12}+ B_{22} \tilde B_{22}   K^{(s)}_{22}\Big| \Big| T^{(s)}_2\Big| \Big|_{*}^2  
=K^{(u)}_{22} \Big| \Big| T^{(u)}_2\Big| \Big|_{*}^2 \,,
\ee
which, by plugging in  equation (\ref{qt1}),  can be written as
\be
\frac{\tilde B_{22}}{B_{11}} \det B K^{(s)}_{22}\Big| \Big| T^{(s)}_{2}\Big| \Big|_{*}^2  =K^{(u)}_{22}\Big| \Big| T^{(u)}_{2}\Big| \Big|_{*}^2 \,.
\ee
The determinant can be evaluated using the  Frobenious formula and gives
\footnote{
It is convenient to first write $A=a_1 b_1$, $B=a_1 b_2$ and
$C=q\frac{a_1}{a_2}$ and then apply  the Frobenius determinant (for example see \cite{spiri}):
\be
{\rm det}_{1\leq i,j \leq N}\left( \frac{\theta_{11}(t^{-1} a_i^{-1} b_j^{-1} ;q)}{\theta_{11}(t^{-1} , a_i^{-1}  b_j^{-1} ;q)}\right)
=\frac{\theta_{11}(t^{-1}  \prod_{i}^N a_i^{-1}  b_i^{-1} ;q)}{\theta_{11}(t^{-1} )}
\frac{\prod_{1\leq i<j \leq N} a_j  b_j  \theta_{11}(a_j/a_i;q) \theta_{11}(b_j/b_i;q)  }{\prod_{1\leq i,j \leq N}   \theta_{11}(a_i^{-1}  b_j^{-1} ;q) \, }\,.
\ee
}

\be
\det B=qBC^{-1}\frac{(q^{-1}C;q)_\infty(qBA^{-1};q)_\infty}{(C;q)_\infty(BA^{-1};q)_\infty}\frac{\theta_{11}(rz^{-1};q)}{\theta_{11}(z^{-1};q)}\,\\
\ee
and 
\ben
\frac{\tilde{B}_{22}}{B_{11}}\det B=
\Big|\Big|\frac{(q^{-1}C;q)_\infty(qBA^{-1};q)_\infty}{(B;q)_\infty(CA^{-1};q)_\infty}
\frac{\theta_{11}(qrz^{-1};q)}{\theta_{11}(qA^{-1}z^{-1};q)}\Big| \Big|_{*}^2\,.
\een
We then obtain the following equation for the  three-point function
\ben
\label{seq}
&&\frac{K^{(u)}_{22}}{K^{(s)}_{22}}=\Big|\Big|\frac{(q^{-1}C;q)_\infty(qBA^{-1};q)_\infty}{(B;q)_\infty(CA^{-1};q)_\infty}
\frac{\theta_{11}(qrz^{-1};q)}{\theta_{11}(qA^{-1}z^{-1};q)}\Big| \Big|_{*}^2 \Big| \Big|\frac{T^{(s)}_{2}}{T^{(u)}_{2}}   \Big| \Big|_{*}^2 \,.
\een
\end{enumerate}
The bootstrap equations  (\ref{feq}) and (\ref{seq}) are the $q$-analog of (\ref{k22}) and (\ref{diag}) derived for Liouville theory.  In the next section we will show that, for two specific pairings of the chiral sectors, the bootstrap equations can be solved and the three-point functions explicitly determined.

\subsection{$id$-pairing three-point functions}
\label{idsec}

In this section we determine the three-point function for the $q$-deformed correlators involving $id$-pairing of conformal blocks.
We begin by recording  the relation between the parameters labeling the primary operators and the parameters appearing in the hypergeometrics. We define the following variables
\ben\nn
X_{A}+\frac{m_A}{2 b_0}&=&\alpha_1+\alpha_3+\alpha_4-\frac{b_0}{2}-Q_0\,,\\ \nn
X_{B}+\frac{m_B}{2 b_0}&=&\alpha_1+\alpha_3-\alpha_4-\frac{b_0}{2}\, ,\\
X_{C}+\frac{m_C}{2 b_0}+\frac{1}{b_0}&=&2\alpha_1-b_0=2\alpha_1-Q_0+1/b_0\,,
\een
with $Q_0=b_0+1/b_0$. 
They are  related to hypergeometrics parameters as
\ben\nn
&&A=e^{\beta X_{A}} q^{m_A/2}\,,
\qquad \tilde{A}=e^{-\beta X_{A}} q^{m_A/2}\, ,\\  
\nn
&&B=e^{\beta X_B} q^{m_B/2}, \qquad \tilde{B}=e^{-\beta X_B} q^{m_B/2}\, ,
\\
&&C=q e^{\beta X_C} q^{m_C/2}\,,
\quad \,\, \tilde{C}=\tilde qe^{-\beta X_C} q^{m_C/2}\, ,
\een
and 
\be
q=e^{\beta/ b_0}\, , \quad \tilde q=q^{-1}\, , \quad \tilde z= \bar z.\ee
The tilded and untilded variables just defined, appear in the  $id$-pairing in the following way
\be\label{ipa}
\Big| \Big| f(A,B,C;z;q) \Big| \Big|_{\rm id} ^2= f(A,B,C;z;q) f(\tilde A,\tilde B,\tilde C;\tilde z;\tilde q)\,.
\ee

Using the $id$-pairing, equation (\ref{feq}) gives
\ben\nn
\label{ineq1}
\frac{K^{(s)}_{22} }{K^{(s)}_{11} }=\Big| \Big| \frac{(A;q)_\infty (B;q)_\infty(q^2 C^{-1};q)_\infty  }{(C;q)_\infty (q AC^{-1};q)_\infty (q BC^{-1};q)_\infty  }
  \Big| \Big|_{\rm id}^2
q^{-m_C} \left(e^{\beta X_A}e^{\beta X_B}e^{-\beta X_C}\right)^\frac{m_C}{2} \left(e^{\beta X_C}\right)^{\frac{m_A+m_B-m_C}{2}}\,,
 \nn\\
  \een
while  equation (\ref{seq}) reduces to
\ben
\label{ineq2}\nn
\frac{K^{(u)}_{22}}{K^{(s)}_{22}}&=&\Big|\Big|\frac{(qC^{-1}A;q)_\infty   (q BA^{-1};q)_\infty}{  (B;q)_\infty  (q^2C^{-1};q)_\infty   }\Big|\Big|^2_{\rm id} \nn \\
&&\times e^{-i \pi  \left(m_A-m_B-m_C\right)} q^{\frac{1}{2} \left(-m_A+m_B+m_C\right)} \zeta _A^{m_A-\frac{m_B}{2}-\frac{m_C}{2}} \left(\zeta_B\zeta _C\right){}^{-\frac{m_A}{2}}
\,.
\een

In the following we will focus on the case with $m_A=m_B=m_C=0.$ 
As it will be clear from the mapping to the $S^2\times_q S^1$ theory that we will work out in section \ref{fmap}, 
this corresponds to the case where all the flavor fluxes of the index are turned off $(r_j=l_k=0)$. However, we will keep generic flux ($n\neq0$) for the FI parameter so that $z=\zeta q^{n/2}$ is a  complex variable which will be identified with the cross ratio. 
We notice that, in this case,
\be
a=b_0 X_A\,, \quad b=b_0 X_B\, , \quad c=b_0X_C+1\,,
\ee
 are the same parameters appearing in the undeformed hypergeometrics in (\ref{parmom}).
Moreover, taking the $\beta \to 0$ limit, (the  Virasoro limit of ${\cal V}ir_{q,t}$)
the $id$-pairing of  $q$-hypergeometrics   reduces to the  undeformed  pairing of hypergeometrics  defined in equation (\ref{unpa}):\footnote{If we  take, in analogy with the undeformed case, primaries with momenta $\alpha=Q_0/2+ i p_{\alpha}$, $p_\alpha\in \mathbb{R}$ and   shift $\alpha_3\to \alpha_3-1/(2b_0)$, the variables $X_A,X_B,X_C$ 
become pure imaginary.}
\be
\lim_{q\to 1}\Big| \Big| \phantom{|}_2\Phi_1(A,B,C;q,z) \Big| \Big|_{id} ^2=\phantom{|}_2F_1(a,b,c;z) \phantom{|}_2F_1(a,b,c;\bar z)=\Big| \Big| \phantom{|}_2F_1(a,b,c;z)  \Big| \Big| ^2\,.
\ee
Since  $\beta$ will be identified  with the  $S^1$ length in $S^2\times_q S^1$,  this is consistent with the fact that in the  $\beta \to 0$ limit, the  index  partition function reduces to the $S^2$ partition function which has been shown to match degenerate Liouville correlators \cite{Doroud:2012xw}.

We will now  make an ansatz for the three-point function that solves equations (\ref{ineq1}) and (\ref{ineq2}) 
 (for $m_A=m_B=m_C=0$).
We take\footnote{\label{pref} There could be a  prefactor $P(\alpha_1,\alpha_2,\alpha_3)$ like in the DOZZ formula. However, we will only be looking at bootstrap equations that involve ratios of three-point functions where, in the undeformed case, prefactors cancel-out  and assume that they do still cancel out in the deformed case. As in the undeformed case we do not expect the gauge theory to reproduce them.}
\be\label{qdozz}
C(\alpha_3,\alpha_2,\alpha_1)=\frac{1}{\Upsilon^\beta(2 \alpha_T-Q_0)}\prod_{i=1}^3 \frac{\Upsilon^\beta(2 \alpha_i)}{\Upsilon^\beta(2\alpha_T-2 \alpha_i)}
\ee
where $2\alpha_T=\alpha_1+\alpha_2+\alpha_3$ and the definition and useful properties of the   $\Upsilon^\beta$ function are collected in  appendix \ref{upsi}. This is the $q$-DOZZ three point function that appeared already in \cite{Kozcaz:2010af,Bao:2011rc}.

In the following, using  that the matrices $K^{(s)}_{ij}$ and $K^{(u)}_{ij}$ are related to the three point functions as in the Liouville case (see (\ref{ks}) and  (\ref{ku})),  we verify that the three-point function (\ref{qdozz}) satisfies the bootstrap equations (\ref{ineq1}) and (\ref{ineq2}).

Using equation (\ref{upp1}) we compute the ratio
\be
\frac{C(\alpha_4,\alpha_3,\alpha_1+b_0/2)}{C(\alpha_4,\alpha_3,\alpha_1-b_0/2)}=\left[q^{3/2}C^{-1}\right]^\infty\Big|\Big|\frac{(A;q)_\infty(B;q)_\infty}{(C;q)_\infty(qAC^{-1};q)_\infty(qBC^{-1};q)_\infty}\Big|\Big|_\text{id}^2\frac{\Upsilon^\beta(2\alpha_1+b_0)}{\Upsilon^\beta(2\alpha_1)}\,,
\ee
while  the other ratio gives\footnote{Notice that there are infinities coming from poles of  $\Upsilon^\beta$ but they cancel out in the ratio.}
\ben
\!\!\!\!\!\!\!\!\!\!\!\!\!\!\!\!\!\!\!\!\!\!\!\!\!\!\!\nn\frac{C_-(\alpha_1)}{C_+(\alpha_1)}&=&\frac{C(Q_0-\alpha_1-b_0/2,-b_0/2,\alpha_1)}{C(Q_0-\alpha_1+b_0/2,-b_0/2,\alpha_1)}=\frac{\Big|\Big|(q^2C^{-1};q)_\infty\Big|\Big|_\text{id}^2}{ \left[q^{3/2}C^{-1}\right]^\infty}\frac{\Upsilon^\beta(-2\alpha_1+Q_0)}{\Upsilon^\beta(-2\alpha_1+Q_0-b_0)}\,.
\een
Putting altogether we get
\be
\label{ratio1}
\frac{C(\alpha_4,\alpha_3,\alpha_1+b_0/2)}{C(\alpha_4,\alpha_3,\alpha_1-b_0/2)}\frac{C_-(\alpha_1)}{C_+(\alpha_1)}=
\frac{K^{(s)}_{22} }{K^{(s)}_{11} }=\Big| \Big| \frac{(A;q)_\infty (B;q)_\infty(q^2 C^{-1};q)_\infty  }{(C;q)_\infty (q AC^{-1};q)_\infty (q BC^{-1};q)_\infty  }  \Big| \Big|_{\rm id}^2\,,
\ee
in agreement with equation (\ref{ineq1}). Similarly for the other ratio we find
\ben
\frac{K^{(u)}_{22}}{K^{(s)}_{22}}&=&\Big|\Big|\frac{(qAC^{-1};q)_\infty(q^{-1}C;q)_\infty}{(B;q)_\infty(AB^{-1};q)_\infty}\Big|\Big|^2_{\rm id}\,,
\een
in agreement with equation (\ref{ineq2}). The bootstrap approach, applied to a $q$-deformation of Liouville where the chiral blocks are glued by the $id$-pairing (\ref{ipa}), has allowed us to compute the three-point function of non-degenerate states.


\subsection{$S$-pairing three-point functions}
\label{ssec}

In this section we determine the three-point function for the $q$-deformed  correlators involving $S$-pairing of conformal blocks.
We begin by specifying how the momenta labelling  the primaries are related to the parameters appearing in the
hypergeometrics. We define
\ben\nonumber
X_{A}&=&\alpha_1+\alpha_3+\alpha_4-\frac{\omega_3}{2}-E\,,\\ \nonumber
X_{B}&=&\alpha_1+\alpha_3-\alpha_4-\frac{\omega_3}{2}\,,\\
X_{C}&=&2\alpha_1-\omega_3\,,
\een
and 
\ben\nn\label{elva}
A&=&e^{2\pi iX_{A}/\omega_2}\quad B=e^{2\pi iX_{B}/\omega_2}\,,\quad C=e^{2\pi iX_{C}/\omega_2}\,,\quad q=e^{2\pi i \frac{\omega_1}{\omega_2}}\,,\quad z=e^{2\pi i  Z /\omega_2}\, ,\\
\tilde{A}&=&e^{2\pi iX_{A}/\omega_1},\quad \tilde{B}=e^{2\pi iX_{B}/\omega_1},\quad \tilde{C}=e^{2\pi iX_{C}/\omega_1}\,,
\quad \tilde q=e^{2\pi i \frac{\omega_2}{\omega_1}}\,,\quad \tilde z=e^{2\pi i Z /\omega_1}\,.\qquad
\een
Also in this case, we assume that the matrices $K^{(s)}_{ij}$ and $K^{(u)}_{ij}$ are related to the three-point functions as in the undeformed case (see (\ref{ks}) and  (\ref{ku})). However,  now the parameter associated to the momenta of the degenerate primary is $\omega_3$, so 
the internal channel  states are given by $\beta_1^{(s)}=\alpha_1-\frac{\omega_3}{2}$, $\beta_2^{(s)}=\alpha_1+\frac{\omega_3}{2}$, $\beta_1^{(u)}=\alpha_4-\frac{\omega_3}{2}$ and $\beta_2^{(u)}=\alpha_4+\frac{\omega_3}{2}$.

The parameters $\omega_1$ and $\omega_2$ will be related in the next section to the squashing parameters of the ellipsoid, so we also introduce $Q=\omega_1+\omega_2$ and assume $\omega_1\cdot\omega_2=1$. As we will desrcibe in section \ref{s5}, the freedom  to permute the way we identify  the $\omega_1,\omega_2,\omega_3$ to the squashing parameters and to the degenerate momentum is related to the fact that we think our ellipsoid as a defect inside a squashed $S^5$.

The $S$-pairing is given by 
\be\label{spa}
\Big| \Big| f(A,B,C;z;q) \Big| \Big|_{S} ^2= f(A,B,C;z;q) f(\tilde A,\tilde B,\tilde C;\tilde z;\tilde q)\,,
\ee
where the variables  that enter in (\ref{spa}) are defined in (\ref{elva}). In these variables we have $\Big| \Big|\theta(A;q)\Big| \Big|_{S}^2= \rm{C}^{-2} e^{-i\pi X_A^2}$. 

Using the $S$-pairing,  equation (\ref{feq}) simplifies to\footnote{Up to factors of $\rm{C}$ which could be reabsorbed by redefining $T_A(x)\to T'_A(x)=\frac{\theta(A  \mu)\theta(\mu)}{\theta(x \mu) \theta(A\mu)}$ 
 since $ \Big| \Big|T_A(z)\Big| \Big|_{S}^2  {\rm C}^{-2}=  \Big| \Big|T'_A(z)\Big| \Big|_{S}^2=e^{-i\pi X^2_A}$.
}
\ben\nn
\label{eli1}
\frac{K^{(s)}_{22} }{K^{(s)}_{11} }=
\Big| \Big| \frac{(A;q)_\infty (B;q)_\infty(q^2 C^{-1};q)_\infty  }{(C;q)_\infty (q AC^{-1};q)_\infty (q BC^{-1};q)_\infty  }  \Big| \Big|_{S}^2e^{-i\pi (X_C-Q) (X_C-X_A-X_B+Q)}\,,\nn\\
\een
while equation (\ref{seq}) yields
\ben\nn
\label{eli2}
\frac{K^{(u)}_{22}}{K^{(s)}_{22}}&=&\Big|\Big|\frac{(q^{-1}C;q)_\infty(qBA^{-1};q)_\infty}{(B;q)_\infty(CA^{-1};q)_\infty}
\frac{\theta_{11}(qrz^{-1};q)}{\theta_{11}(qA^{-1}z^{-1};q)}\Big| \Big|_{S}^2 \Big| \Big|\frac{T^{(s)}_{2}}{T^{(u)}_{2}}   \Big| \Big|_{S}^2 =\\\nn
&&=\Big|\Big|\frac{(qC^{-1}A;q)_\infty(qBA^{-1};q)_\infty}{(B;q)_\infty(q^2C^{-1};q)_\infty}
\frac{\theta_{11}(qC^{-1};q)}{\theta_{11}(A C^{-1};q)}
\frac{\theta_{11}(qrz^{-1};q)}{\theta_{11}(qA^{-1}z^{-1};q)}\Big| \Big|_{S}^2 \Big| \Big|\frac{T^{(s)}_{2}}{T^{(u)}_{2}}   \Big| \Big|_{S}^2 =\\
&&=
\Big|\Big|\frac{(qC^{-1}A;q)_\infty   (q BA^{-1};q)_\infty}{  (B;q)_\infty  (q^2C^{-1};q)_\infty   }\Big|\Big|^2_S
e^{-i\pi\left(Q-X_A\right)\left(Q+X_A-X_B-X_C\right) }\,.
\een
Equations  (\ref{eli1}) and (\ref{eli2}) can be used to determine the three-point functions for the $S$-pairing. 
We consider the following ansatz for the three-point function\footnote{See footnote \ref{pref}.}
\be
\label{4p1l}
C(\alpha_3,\alpha_2,\alpha_1)=\frac{1}{S_3(2 \alpha_T-E)}\prod_{i=1}^3 \frac{S_3(2 \alpha_i)}{S_3(2\alpha_T-2 \alpha_i)}\,,
\ee
where $2\alpha_T=\alpha_1+\alpha_2+\alpha_3$ and $E=\omega_1+\omega_2+\omega_3$. The definition and several properties of the triple-sine function $S_3(X)$ are collected in the appendix \ref{3sap}.

We now use this ansatz, equation (\ref{s2fac}), and the property  $S_3(X)=S_3(E-X)$ to compute the three-point functions ratios that appear on the LHS of (\ref{eli1}) and (\ref{eli2}). 
We find
\ben
\nn\frac{K^{(s)}_{22}}{K^{(s)}_{11}}&=&
\frac{S_2(2Q-\alpha_1+\omega_3) S_2(\alpha_1+\alpha_3+\alpha_4-\omega_3/2-E)
S_2(\alpha_1+\alpha_3-\alpha_4-\omega_3/2)}{S_2(2\alpha_1-\omega_3)S_2(\alpha_3+\alpha_4-\alpha_1-\omega_3/2)S_2(Q+\alpha_3-\alpha_4-\alpha_1+\omega_3/2)}=
\\
&&=\frac{S_2(2Q-X_C)S_2(X_{A})S_2(X_B)}{S_2(X_C)S_2(Q+X_{A}-X_C)S_2(Q+X_{B}-X_C)}=
\nn\\
&&=\Big| \Big| \frac{(A;q)_\infty (B;q)_\infty(q^2 C^{-1};q)_\infty  }{(C;q)_\infty (q AC^{-1};q)_\infty (q BC^{-1};q)_\infty  }  \Big| \Big|_{S}^2e^{i\pi  (Q-X_C)(Q+X_C-X_A-X_B)}\,,
\een
in agreement with equation (\ref{eli1}).  The other ratio yields:
\ben
\nn&&\frac{K^{(u)}_{22}}{K^{(s)}_{22}}=\frac{S_2(\alpha_3+\alpha_4-\alpha_1-\omega_3/2) S_2(2Q-2\alpha_4+\omega_3)}{
S_2(\alpha_3+\alpha_1-\alpha_4-\omega_3/2) S_2(2Q-2\alpha_1+\omega_3)}=\\\nn
&&=
\frac{S_2(X_{A}-X_{C}+Q) S_2(Q+X_{B}-X_{A})}{
S_2(X_{B}) S_2(2Q-X_{C})}=\\&&=
\Big|\Big|\frac{(qC^{-1}A;q)_\infty   (q BA^{-1};q)_\infty}{  (B;q)_\infty  (q^2C^{-1};q)_\infty   }\Big|\Big|^2_S
e^{-i\pi\left(Q-X_A\right)\left(Q+X_A-X_B-X_C\right) }\,,
\een
in agreement with equation  (\ref{eli2}).  This shows that the three point function defined in (\ref{4p1l}) solves the equations imposed by the bootstrap method, confirming   the exactness of (\ref{4p1l}).

\section{3d partition functions as  $q$-deformed CFT correlators}\label{fmap}

In this section we map $q$-deformed CFT degenerate correlators to 3d gauge theory partition functions
working out a dictionary between parameters.
We begin by rewriting the equation expressing the flop invariance of gauge theory partition functions--as in
equations (\ref{flopin}) and  (\ref{flopel})--dividing each side by $G^{(1),I}_{cl}\cdot G^{(1),I}_{1loop} $
\ben
\!\!\!\!\Big| \Big| \mathcal{Z}^{(1),I}_{V}\Big| \Big|_{*}^2
+\frac{G^{(2),I}_{cl}\cdot G^{(2),I}_{1loop} }{G^{(1),I}_{cl}\cdot G^{(1),I}_{1loop} } \cdot\Big| \Big| \mathcal{Z}^{(2),I}_{V}\Big| \Big|_{*}^2
=
\frac{G^{(1),II}_{cl}\cdot G^{(1),II}_{1loop} }{G^{(1),I}_{cl}\cdot G^{(1),I}_{1loop} } 
\cdot\Big| \Big| \mathcal{Z}^{(1),II}_{V}\Big| \Big|_{*}^2
+\frac{G^{(2),II}_{cl}\cdot G^{(2),II}_{1loop} }{G^{(1),I}_{cl}\cdot G^{(1),I}_{1loop} } \cdot\Big| \Big| \mathcal{Z}^{(2),II}_{V}\Big| \Big|_{*}^2 \, \nonumber\\
\een
and  equation (\ref{qcs}), expressing crossing symmetry in CFT  dividing each side by $K_{11}^{(s)} $
 \be
\Big| \Big| I^{(s)}_1  \Big| \Big|_{*}^2+
\frac{K_{22}^{(s)}}{K_{11}^{(s)} }
\Big| \Big| I^{(s)}_2  \Big| \Big|_{*}^2=
\frac{K_{11}^{(u)}}{K_{11}^{(s)} }
\Big| \Big| I^{(u)}_1  \Big| \Big|_{*}^2+
\frac{K_{22}^{(u)}}{K_{11}^{(s)} }
\Big| \Big| I^{(u)}_2  \Big| \Big|_{*}^2\,.
\ee

The identification of quantities in the first channel gives:
\be
\label{map1}
\mathcal{Z}^{(1),I}_{V}= F^{(s)}_1(z)\,, \qquad \mathcal{Z}^{(2),I}_{V}= F^{(s)}_2(z)\,,
\ee
\be
\label{map2}
\frac{G^{(2),I}_{cl} }{G^{(1),I}_{cl} } =\Big| \Big|
T^{(s)}_2\Big| \Big|_{*}^2\,,
\ee
\be
\label{map3}
\frac{K^{(s)}_{22} }{K^{(s)}_{11} }=\frac{G^{(2),I}_{1loop} }{G^{(1),I}_{1loop} } \,.
\ee

Similarly,  the second channel  yields:

\be
\label{map4}
\mathcal{Z}^{(1),II}_{V}= F^{(u)}_1(z)\,  , \qquad \mathcal{Z}^{(2),II}_{V}= F^{(u)}_2(z)\,,
\ee

\be
\label{map5}
\frac{G^{(1),II}_{cl} }{G^{(1),I}_{cl} } =\Big| \Big|
T^{(u)}_1\Big| \Big|_{*}^2\, , \qquad \frac{G^{(2),II}_{cl} }{G^{(1),I}_{cl} } =\Big| \Big|
T^{(u)}_2\Big| \Big|_{*}^2\, ,
\ee
and
\be
\label{map6}
\frac{K^{(u)}_{11} }{K^{(s)}_{11} }=\frac{G^{(1),II}_{1loop} }{G^{(1),I}_{1loop} } \, , \qquad  \frac{K^{(u)}_{22} }{K^{(s)}_{11} }=\frac{G^{(2),II}_{1loop} }{G^{(1),I}_{1loop} } \,.
\ee
To match  vortex partition functions and conformal blocks in equation (\ref{map1}), we need to identify 
the parameters $A,B,C$ of the hypergeometric functions appearing in the gauge theory and in the CFT correlators.
Introducing $\phi_i=e^{i \beta\, \Phi_i}$, $\xi_i=e^{i\beta\, \Xi_i}$, we obtain the following dictionary:

\be
\begin{array}{|c|c|}
\hline
\text{Ellipsoid}&\text{Index}\\
\hline
X_A=-i(\tilde{m_1}-m_1)&X_A=i(\Phi_1-\Xi_1),\quad m_A=r_1-l_1\\
&\\
X_B=-i(\tilde{m_2}-m_1)&X_B=i(\Phi_1-\Xi_2),\quad m_B=r_1-l_2\\
&\\
X_C=-i(m_2-m_1+iQ)&X_C=i(\Phi_1-\Phi_2),\quad m_C=r_1-r_2\\
&\\
\omega_1=\omega_2^{-1}=b,\,& \beta=\text{lenght of}\,\, S^1\\ 
&\\
\alpha_1=\frac{E}{2}+i\frac{m_1-m_2}{2}&\alpha_1=\frac{Q_0}{2}+(r_1-r_2)\frac{\omega_2}{4}+i\frac{\Phi_1-\Phi_2}{2}\\
&\\
\alpha_3=\frac{\omega_3}{2}-i\frac{\tilde m_1+\tilde m_2-m_1-m_2}{2} &\hat{\alpha}_3=\frac{\omega_1}{2}+(r_1+r_2-l_1-l_2)\frac{\omega_2}{4}-i\frac{\Xi_1+\Xi_2-\Phi_1-\Phi_2}{2}\\
&\\
\alpha_4=\frac{E}{2}-i\frac{\tilde m_1-\tilde m_2}{2}&\alpha_4=\frac{Q_0}{2}+(l_2-l_1)\frac{\omega_2}{4}-i\frac{\Xi_1-\Xi_2}{2}\\
&\\
\hline
\multicolumn{2}{|c|}{z_\text{CFT}=qr^{1/2}z^{-1}_\text{gauge}}\\
\hline
\end{array}
\ee

where for the index we shifted $\alpha_3\rightarrow \hat{\alpha}_3=\alpha_3-\omega_2/2$ and for the ellipsoid we defined $ Q=b+1/b$.

With this dictionary it is then easy to check all other equations (\ref{map2}),(\ref{map3}),(\ref{map4}),(\ref{map5}),(\ref{map6}).
For example, equation (\ref{map3}) for the ellipsoid gives
\ben
\frac{K^{(s)}_{22} }{K^{(s)}_{11} }&=&\frac{S_2(2Q-X_C)S_2(X_{A})S_2(X_B)}{S_2(X_C)S_2(Q+X_{A}-X_C)S_2(Q+X_{B}-X_C)}=\nn\\
&=&\frac{s_b(m_1-m_2+iQ/2)  s_b(\tilde m_1-m_1-iQ/2)s_b(\tilde m_2-m_1-iQ/2)}{s_b(\tilde m_1-m_2-iQ/2)s_b(\tilde m_2-m_2-iQ/2) s_b(m_2-m_1+iQ/2) }=\frac{G^{(2),I}_{1loop} }{G^{(1),I}_{1loop} }\, .\nn\\
\een
While for the index  (\ref{map3}) gives
\ben\
\frac{K^{(s)}_{22} }{K^{(s)}_{11} }&=&\Big| \Big| \frac{(A;q)_\infty (B;q)_\infty(q^2 C^{-1};q)_\infty  }
{(C;q)_\infty (q AC^{-1};q)_\infty (q BC^{-1};q)_\infty  }
\Big| \Big|_{id}^2 q^{-m_C} \left(e^{\beta X_A}e^{\beta X_B} e^{-\beta X_C}\right)^{\frac{m_C}{2}} \left(e^{\beta X_C}\right)^\frac{m_A+m_B-m_C}{2}=\nn\\
&=&\Big|\Big|\frac{(q\phi_2\phi_1^{-1}q^\frac{r_2-r_1}{2})(\phi_1\xi_1^{-1}q^\frac{r_1-l_1}{2})(\phi_1\xi_2^{-1}q^\frac{r_1-l_2}{2})}
{(q\phi_1\phi_2^{-1}q^\frac{r_1-r_2}{2})(\phi_2\xi_1^{-1}q^\frac{r_2-l_1}{2})(\phi_2\xi_2^{-1}q^\frac{r_2-l_2}{2})}\Big|\Big|^2_{id}\times\nn\\
&&\times
q^{r_2-r_1} \left(\xi _1^{-1} \xi _2^{-1}\phi_1\phi_2\right)^\frac{r_1-r_2}{2} (\phi _1\phi_2^{-1})^\frac{r_1+r_2-l_1-l_2}{2}=\frac{G^{(2),I}_{1loop} }{G^{(1),I}_{1loop} }
\een
 where the first equality follows from (\ref{ineq1}) and  agrees with the three-point function computation for zero fluxes.

The flop symmetry described in formulas (\ref{flopin}) and  (\ref{flopel}), is therefore realized by the crossing symmetry of the $q$-deformed correlators we have constructed in the previous section.

\section{5d partition functions as $q$-deformed CFT correlators}
In this section we argue that partition functions of 5d ${\cal N}=1$ gauge theories on  $S^5$ and $S^4\times  S^1$ can be mapped to $q$-deformed CFT correlation functions of non-degenerate states. We start looking  at the $S^4\times  S^1$ partition function, \emph{i.e.} the superconformal index.  

\subsection{Partition function on $S^4\times S^1$ and $q$-deformed CFT}\label{5dind}

The partition function for 5d ${\cal N}=1$ supersymmetric gauge theories on $S^4\times S^1$ has been computed in \cite{Kim:2012gu, Terashima:2012ra}  using a localization scheme similar to the one developed in \cite{Pestun:2007rz} for gauge theory on  $S^4$. A derivation that uses topological strings is given in \cite{Iqbal:2012xm}. The result is written in terms of an integral over the constant value of the vector potential along the $S^1$ direction  $A_\tau=\sigma$,  and the integrand has a perturbative  contribution ${\cal Z}_{\text{1-loop}}$ and an instanton contribution ${\cal Z}_{\text{inst}}$.  In details 
\be\label{s4s1}
Z_{S^4\times S^1}=\int d \sigma {\cal Z}_{\text{1-loop}}(\sigma)  {\cal Z}_{\text{inst}}(\sigma)\,,
\ee
where the integration measure is included in the perturbative part and the explicit expression of the various factors depend on the field content of the theory. For the one-loop part, it results that the vector multiplet gives the following contribution
\be
{\cal Z}^{\text{vect}}_{\text{1-loop}}(\sigma)=\prod_{\alpha>0}\Upsilon^\beta\left(i\alpha(\sigma)\right)\Upsilon^\beta\left(-i\alpha(\sigma)\right)\,,
\ee
where $\alpha$ are the roots of the gauge group and we have rewritten the results of \cite{Kim:2012gu, Terashima:2012ra} in terms of  the function  $\Upsilon^\beta(X)$  defined in  (\ref{upsi}),   using a relation between $\Upsilon^\beta(X)$  and  the standard $\Upsilon(X)$ given in (\ref{relupsi}).   A hypermultiplet of mass $m$ in a representation $R$ of the gauge group contributes as
\be
{\cal Z}^{\text{hyper}}_{\text{1-loop}}(\sigma,m,R)=\prod_{\rho\in R}\Upsilon^\beta\left(i(\rho(\sigma)+m)+\frac{Q_0}{2}\right)^{-1}\,,
\ee
where $\rho$ are the weights of the representation $R$.  Like for the $S^2\times S^1$ case, $\beta$ is the period of the compact direction $S^1$. The radius of the $S^4$ is equal to the unity and $Q_0=b_0+1/b_0$, where $b_0$ can be related to the squashing parameter of the $S^4$, as in  \cite{Hama:2012bg}.

The instanton contribution is due to point-like instantons localized at the poles of the $S^4$, where the metric is effectively ${\mathbb R}^4\times S^1$. It follows that the contribution from each of the two poles is  given by  the Nekrasov function for 5d gauge theory compactified on a circle ${\cal Z}^{5d}_{\text{Nek}}$ \cite{Nekrasov:2002qd, Nekrasov:2003rj}.  Instantons localized at the north and south poles come with opposite  topological charge,  therefore the total contribution is given by  ${\cal Z}_{\text{inst}}={\cal Z}^{5d}_{\text{Nek}}\,\bar{{\cal Z}}^{5d}_{\text{Nek}}=|{\cal Z}^{5d}_{\text{Nek}}|^2$, similar to the $S^4$ case \cite{Pestun:2007rz}.

We now argue that the partition function (\ref{s4s1}) can be mapped to a non-degenerate $q$-deformed CFT correlator.  This relation is a natural generalization of the AGT correspondence \cite{Alday:2009aq}, where partition functions of gauge theories on $S^4$ are mapped to correlation functions of Liouville/Toda theory.
The mapping between $q$-deformed Virasoro/W algebra conformal blocks and  5d instanton partition functions has been already discussed in the literature \cite{Awata:2009ur,Awata:2010yy,Schiappa:2009cc,Mironov:2011dk,Yanagida:2010vz}. 
As the instanton contribution is given by a complex modulus squared, we consider a correlator where the two chiral sectors are glued by the $id$-pairing (\ref{ipa}). Indeed, when the flavor fluxes  are switched off, the tilded and untilded variables of the $id$-pairing are related by complex conjugation.  
Another motivation for using the $id$-pairing comes from the interpretation of the 3d index theory as  a defect theory of the  5d index  theory.  In analogy with \cite{Alday:2009fs}, we would then associate the defect theory to 
a correlation function involving degenerate primaries.

Let us  consider the case where the 5d gauge theory is an $SU(2)$ vector multiplet coupled to  four fundamental hypermultiplets.  Like in the AGT case, we propose that the partition function of  this gauge theory is equivalent to the correlation function for four non-degenerate states, where the two chiral blocks are glued using the   $id$-pairing described in section \ref{idsec}.  In analogy to the AGT correspondence, the integration over the zero mode $\sigma$ is mapped to the integration over the states in the internal channel of the conformal blocks,  the total one-loop factor is mapped to the product of three point functions and the north and south pole instantons function are mapped to the holomorphic and anti-holomorphic conformal blocks. 

The equivalence between the four point conformal block and the related 5d instanton function was discussed in \cite{Awata:2010yy,Mironov:2011dk}. In particular, the  dictionary between the parameters of the ${\cal V}ir_{q,t}$ algebra and the equivariant parameters worked out in \cite{Awata:2009ur,Awata:2010yy} gives $q=e^\frac{\beta}{b_0},t=e^{-\beta b_0}$. Therefore the parameters  satisfy the relation $t=q^{-b_0^2}$ as in (\ref{lvir}), and in the limit $\beta\rightarrow 0$ the ${\cal V}ir_{q,t}$ becomes the Virasoro algebra with central charge $c_V=1+6Q_0^2$. This is in agreement with the fact that in the same limit, the  $S^4\times S^1$   theory  reduces to the  $S^4$ theory that is related to the Virasoro algebra.

In the following we show that the three-point function for  $id$-pairing (\ref{qdozz}) reproduces the one-loop factor.
Considering the dictionary\footnote{ $\sigma$ takes value in the Cartan, therefore $\sigma=(\tilde\sigma,-\tilde\sigma)$. In the following we will rename $\tilde\sigma\rightarrow\sigma$.}
\be
\alpha=i\sigma+\frac{Q_0}{2}\,,\quad\alpha_1+\alpha_2=im_1+Q_0\, , \quad \alpha_1-\alpha_2=im_2\,,\quad\alpha_3+\alpha_4=im_3+Q_0\, , \quad \alpha_3-\alpha_4=im_4\,,
\ee
it follows that\footnote{We use $\Upsilon^\beta(X)=\Upsilon^\beta(Q_0-X)$.}
\ben
C(\alpha_1,\alpha_2,\alpha)C(Q_0-\alpha,\alpha_3,\alpha_4)&=&
{\cal Z}^{\text{vect}}_{\text{1-loop}}(\sigma)\prod_{i=1}^4{\cal Z}^{\text{hyper}}_{\text{1-loop}}(\sigma,m_i,F)\,,
\een
that is the total one-loop contribution of the  gauge theory with four hypermultiplets in the fundamental representation $F$.
\footnote{As in the AGT case, the equality is up to factors independent on $\sigma$.}

\subsection{Partition function on squashed $S^5$ and $q$-deformed CFT}\label{s5}
The partition function of 5d ${\cal N}=1$ supersymmetric gauge theory on the squashed $S^5$ has been studied in \cite{Imamura:2012bm, Lockhart:2012vp, Kim:2012qf}, extending previous results for the round $S^5$ \cite{Kallen:2012cs, Hosomichi:2012ek, Kallen:2012va, Kim:2012ava}.  The path integral reduces to the integral over the saddle points, that are characterized by string instantons winding an $S^1$ fiber of the Hopf fibration of the five sphere,  and a zero mode of an adjoint scalar in the vector multiplet, \emph{i.e.} a matrix $\sigma$.  The  result is therefore written in terms of a matrix model. The integrand is given by  a classical part ${\cal Z}_{\text{cl}}$, a one-loop factor ${\cal Z}_{\text{1-loop}}$ and a non-perturbative contribution ${\cal Z}_{\text{inst}}$ that arises from integrating over the moduli space of the instantons. The partition function is given by 
\be\label{s5}
Z_{S^5}=\int d \sigma {\cal Z}_{\text{cl}}(\sigma) {\cal Z}_{\text{1-loop}}(\sigma)  {\cal Z}_{\text{inst}}(\sigma) 
\ee
where the integration measure has been inserted in the one-loop part. The three factors in the integrand have expressions  that depend on the field content of the theory \cite{Imamura:2012bm, Lockhart:2012vp, Kim:2012qf}. The instanton contribution assumes a factorized form \cite{Lockhart:2012vp, Kim:2012qf}
\be
{\cal Z}_{\text{inst}}={\cal Z}_{\text{inst}}^{1}{\cal Z}_{\text{inst}}^{2}{\cal Z}_{\text{inst}}^{3}\,.
\ee
Each factor corresponds to a Nekrasov instanton function on ${\mathbb R}^4\times S^1$  and it is 
associated to one  of the three fixed points of the $\mathbb{CP}^2$ base of the Hopf fibration.
More explicitly, as explained in \cite{Lockhart:2012vp},  we can view $S^5$ as a $T^3$ fibration over a triangle. In the interior of the triangle all the circles are non-vanishing, while on each edge, one of the circles vanishes and finally at  each vertex only one cycle (the $i$-th cycle)  survives.
Notice that since  each edge is a $T^2$ fibration over an interval, we have three 
squashed three-sphere inside $S^5$. 
To each vertex we can associate the following combination of  equivariant parameters
\ben
v_1: \quad \left(1,\frac{\omega_2}{\omega_1},-\frac{\omega_3}{\omega_1}\right)\, ,\qquad 
v_2: \quad \left(\frac{\omega_1}{\omega_2},1,-\frac{\omega_3}{\omega_2}\right)\, ,\qquad
v_3: \quad \left(\frac{\omega_1}{\omega_3},\frac{\omega_2}{\omega_3},1\right)\,.
\een
Hence each vertex $v_i$ contributes to the partition function with a copy of the instanton partition function ${\cal Z}_{\text{inst}}^{i}$ with equivariant parameters $q,t$ given respectively by
\be
\label{qtmap}
(q,t) = \quad  \left(e^{2\pi i \frac{\omega_2}{\omega_1}}, e^{-2\pi i\frac{\omega_3}{\omega_1}} \right)\, ,\quad 
\left(e^{2\pi i \frac{\omega_1}{\omega_2}}, e^{-2\pi i\frac{\omega_3}{\omega_2}} \right)\, ,\quad \left(e^{2\pi i \frac{\omega_1}{\omega_3}}, e^{2\pi i\frac{\omega_2}{\omega_3}} \right)\,.
\ee

The one-loop factor receives the following contribution from the vector multiplet 
\be
{\cal Z}^{\text{vect}}_{\text{1-loop}}(\sigma)=\prod_{\alpha>0} S_3(i\alpha(\sigma)) S_3(-i\alpha(\sigma))\,,
\ee
where $S_3(X)$ is the triple-sine function described in appendix \ref{3sap}. A hypermultiplet  in a representation $R$ and mass $m$ contribute as
\be
{\cal Z}^{\text{hyper}}_{\text{1-loop}}(\sigma,m,R)=\prod_{\rho\in R} S_3\left(i(\rho(\sigma)+m)\frac{E}{2}\right)^{-1}\,,
\ee
where $E=\omega_1+\omega_2+\omega_3$ and   $\omega_1,\omega_2,\omega_3$ are related to the squashing parameters. The round  $S^5$ is obtained setting $\omega_1=\omega_2=\omega_3=1$.  

We propose that the squashed $S^5$ partition  function (\ref{s5}) is related to a correlation function of non-degenerate states. In analogy with the AGT correspondence \cite{Alday:2009aq}, we expect the non-perturbative part to be mapped to the conformal blocks while the one-loop factors to the three-point function contribution.
Since the  instanton part contains three copies of the Nekrasov partition functions on ${\mathbb R}^4\times S^1$, we are lead to consider three copies of  $\mathcal{V}ir_{qt}$ with
$(q,t)$ as in (\ref{qtmap}). As we already mentioned, there are three ellipsoids inside $S^5$ which we  think  as defects theories as in \cite{Lockhart:2012vp}. Hence, in the spirit of \cite{Alday:2009fs}, where  defects are realised in the CFT by degenerate primaries, we  interpret  the $S$-correlators of section \ref{ssec} as degenerate correlators corresponding to ellipsoid defects inside $S^5$.
This in turn suggests that the three-point function for $S$-correlators should be
able to reconstruct the  one-loop part on $S^5$. In what follows we show that this is indeed the case leaving for a future publication \cite{workinprogress}, the study of the instanton sector.

Indeed it is immediate to  show that  the three point function factor of a four-point correlator (calculated using the $S$-pairing three point function defined in (\ref{4p1l})) can be mapped to the one-loop contribution  of a gauge theory with one vector multiplet in the adjoint representation of the gauge group and four  fundamental hypermultiplets. Namely:
\ben
\nonumber C(\alpha_1,\alpha_2,\alpha)C(E-\alpha,\alpha_3,\alpha_4)&=&
{\cal Z}^{\text{vect}}_{\text{1-loop}}(\sigma)\prod_{i=1}^4{\cal Z}^{\text{hyper}}_{\text{1-loop}}(\sigma,m_i,F)\,,
\een
with the  following dictionary
\be
\alpha= i\sigma+\frac{E}{2}\,,\quad\alpha_1+\alpha_2=im_1+E\, , \quad \alpha_1-\alpha_2= im_2\,,\quad \alpha_3+\alpha_4= im_3+E\, , \quad \alpha_3-\alpha_4=im_4\,.
\ee

\section*{Acknowledgments}
We would like to thank G.~Bonelli and  N.~Drukker for discussions.
The work of F.~Nieri is partially  supported  by the EPSRC - EP/K503186/1.

  \appendix

 \section{Index factorization}\label{iid}
In this appendix we  evaluate the integral (\ref{cind}) taking the contribution of poles located at
 \ben
t= \phi_i^{-1} q^{(s+r_i)/2} q^{-k} , && \quad k\geq min(0,s+r_i), \qquad i=1,\cdots N_f
\, . \een
At fixed $i$, we define  $M=s+r_i$.
Below we list the residues at the poles:

 \begin{itemize}

 \item Fundamentals tetrahedra  numerators: 
 \ben
\prod_{l=0}^\infty (1-q^{l+1} t^{-1} \phi_j^{-1}  q^{-(s+r_j)/2})\to \frac{(q x_i x_j^{-1};q)_\infty}{(q x_i x_j^{-1};q)_{k-M}}\, ,
 \een
where we used that
\be
(z q^n;q)_\infty=\frac{(z;q)_\infty}{(z;q)_n}\, .
\ee 
 \item Fundamentals tetrahedra  denominators with ($j\neq i$): 
 \ben
\prod_{l=0}^\infty (1-q^{l} t \phi_j  q^{-(s+r_j)/2})\to ( \tilde x_i \tilde x_j^{-1}; q)_\infty ( \tilde q \tilde x_i \tilde x_j^{-1};\tilde q)_{k}\, , \een
where we used that
 \be
 \prod_{j=0}^\infty (1- a q^{j-k})= (a,q)_\infty  \prod_{j=0}^{k-1}(1-a  q^{-j-1}    )=(a,q)_\infty \prod_{j=0}^{k-1}(1-a  \tilde q^{j+1}    )\, .
 \ee

\item Anti-fundamentals tetrahedra numerators: 
 \ben
 \nonumber
 &&\prod_{l=0}^\infty (1-q^{l+1} t \xi_j   q^{(s+l_j)/2})\to 
 (q y_j x_i^{-1};q)_\infty  ( x_i y_j^{-1};q)_{k-M} (-1)^{k-M}\\
 &&
 q^{-(k-M)(k-M+1 )/2}  (q y_j x_i^{-1})^{k-M}
 \, ,\\
 \een
where we used that
 \be
  \prod_{j=0}^\infty (1- a q^{j-k})= (a,q)_\infty (-a)^k q^{-k(k+1)/2}\prod_{j=0}^{k-1}(1- a^{-1} q^{j+1}    )\,.
 \ee
 
 \item Anti-fundamentals tetrahedra  denominators: 
 \ben
&&\prod_{l=0}^\infty (1-q^{l} t^{-1} \xi_j^{-1}  q^{(s+l_j)/2})\to\prod_{l=0}^\infty 
 (1-  \tilde y_j \tilde x_i^{-1} q^{l+k})=\\
 &&
=\frac{(   \tilde y_j \tilde x_i^{-1};q)_\infty }{(   \tilde y_j \tilde x_i^{-1};q)_k}=( \tilde y_j \tilde x_i^{-1}; q)_\infty 
\frac{(-1)^k q^{-k(k-1)/2}  ( \tilde y_j \tilde x_i^{-1})^{-k}  }{
( \tilde y_j^{-1} \tilde x_i;\tilde q)_k  } \,,
\een

where we used that
 \be
 (A,q)_k=(-1)^k A^k q^{k(k-1)/2}  (A^{-1};\tilde q)_k\, .
 \ee
\end{itemize}
So far we have
\ben
\label{t1}\nn
&&(-1)^M q^{-k} q^{k-M}  q^{M(k-M+1 )/2} q^{kM/2} 
\times\frac{(q x_i x_j^{-1};q)_\infty}{( \tilde x_i \tilde x_j^{-1};q)_\infty}
 \frac{ (q  y_j x_i^{-1};q)_\infty }{( \tilde y_j \tilde x_i^{-1};q)_\infty}
\\&& 
\times\frac{ (x_i y_j^{-1};q)_{k-M}}{(q x_i x_j^{-1};q)_{k-M}}
 (  y_j x_i^{-1})^{k-M}  \times
\frac{( \tilde y_j^{-1} \tilde x_i;\tilde q)_k  }{( \tilde q \tilde x_i \tilde x_j^{-1};\tilde q)_{k}}  ( \tilde y_j \tilde x_i^{-1})^{k}\,.
\een
We now introduce the variables
\ben
k=t\,, \qquad k-M=p\,,
\een
so that 
\be
M=t-p\,, \qquad s=-r_i+t-p\,.
\ee
We can therefore write (\ref{t1}) as
\ben
\label{h1}
\nn&&((-1)^{t-p}q^{(t^2-p^2)/2}  q^{(p-t)/2} )^{N_f}
\times\frac{(q x_i x_j^{-1};q)_\infty}{( \tilde x_i \tilde x_j^{-1};q)_\infty}
 \frac{ (q  y_j x_i^{-1};q)_\infty }{(  \tilde y_j \tilde x_i^{-1};q)_\infty}
\\&& 
\times\frac{ ( x_i y_j^{-1};q)_{p}}{(q x_i x_j^{-1};q)_{p}}
 (  y_j x_i^{-1})^{p}  \times
\frac{( \tilde y_j^{-1} \tilde x_i;\tilde q)_t  }{( \tilde q \tilde x_i \tilde x_j^{-1};\tilde q)_{t}}  (\tilde y_j \tilde x_i^{-1})^{t}\,.
\een
We still need to compute the contribution of the prefactors in the definition of $\chi$ (see equation (\ref{chi}))
and the classical term:
\begin{itemize}
\item The classical term contributes as
\ben
\label{h2}
\nn
&&t^n \omega^s \to \omega^s ( \phi_i^{-1}  q^{(s+r_i)/2} q^{-k})^n=
\omega^s ( \phi_i^{-1})^n  q^{-n(k-M)/2} q^{-n k/2}= \\
&&=\omega^{-r_i+t-p} (  \phi_i^{-1})^n  q^{-n p/2} q^{-n t/2}=
\omega^{-r_i}(  \phi_i^{-1})^n w^{-p} \tilde w^{-t}
 \,,
\een
where we also defined
\be
z=\omega q^{n/2}\, , \quad \tilde z=\omega^{-1} q^{n/2}\,.
\ee
\item The prefactors in (\ref{chi}) give

\ben
\label{h3}
\nn
&&q^{-(s+r_j)/4} (t \phi_j)^{(s+r_j)/2} q^{(s+l_k)/4}  (t \xi_k)^{(s+l_k)/2}|_{t=\phi_i^{-1}q^{-k}  q^{(s+r_i)/2}}\\
&&=( x_i (y_k x_j)^{-1/2} )^{p} ( \tilde x_i (\tilde y_k \tilde x_j)^{-1/2} )^{t} q^{N_f(p^2-t^2)/2} 
 \nn \\
&&\times \phi_i^{r_i}\phi_i^{- (r_j+l_k)/2 }  ( \phi_j^{1/2}  \xi_k^{1/2})^{-r_i}  \times
  \phi_j^{r_j/2}
\xi_k^{l_k/2}q^{-(r_j-l_k)/4}\,.   
\een
\end{itemize}

Combining the infinite products in (\ref{h1}) with the second line in (\ref{h3}) we find
\ben
&&\frac{(q x_i x_j^{-1};q)_\infty}{( \tilde x_i \tilde x_j^{-1};q)_\infty}
 \frac{ (q  y_j x_i^{-1};q)_\infty }{(  \tilde y_j \tilde x_i^{-1};q)_\infty} \phi_i^{r_i}\phi_i^{- (r_j+l_k)/2 }  ( \phi_j^{1/2}  \xi_k^{1/2})^{-r_i}   \phi_j^{r_j/2}
\xi_k^{l_k/2}q^{-(r_j-l_k)/4}=\nn\\
&&=
\Big|\Big| (q x_i x_j^{-1};q)_\infty  (q  y_j x_i^{-1};q)_\infty  \Big|\Big|^2_{\rm id} (q^{1/2} \phi_i \phi_j^{-1})^{(r_i-r_j)/2}
(q^{1/2} \xi_k \phi_i^{-1})^{(l_k-r_i)/2}
 =\\
&&=
\chi(\phi_j \phi_i^{-1},r_j-r_i)  \chi( \phi_i \xi_k^{-1},r_i-l_k):=  G^{(i)}_{1loop} \,.
\een

Combining what is left in (\ref{h1}) with the first line in (\ref{h3}) and the part depending on $p$ and $t$ in (\ref{h2}) we find
\be
\sum_p \prod_{j,k}^{N_f} \frac{ ( x_i y_k^{-1};q)_{p}}{(q x_i x_j^{-1};q)_{p}}
 ( (-q^{1/2})^{N_f } z^{-1}   \prod_{j,k}^{N_f}  y_k^{1/2} x_j^{-1/2})^{p}  :=Z_v^{i}\,.
\ee
Finally what is left in equation (\ref{h2}) gives
\be
\omega^{-r_i}(  \phi_i^{-1})^n:=G^{(i)}_{cl}\,.
\ee

\section{Special functions}
We describe in this appendix few  special functions used in the main text.

\subsection{$r$-gamma functions and  $r$-sine functions}\label{3sap}
The $r$-gamma function can be defined as the following regularized infinite product
\be
\Gamma_r(z|\vec{E})\sim \prod_{n_1,\cdots ,n_r=0}^{+\infty}(\vec E \cdot \vec n +z)^{-1}\,,
\ee
where $\vec E=\omega_1,\ldots ,\omega_r$ and $\vec n=n_1,\ldots ,n_r$. The $r$-sine function is defined as
\be
S_r(z|\vec{E})\sim\Gamma_r(z|\vec{E})^{-1} \Gamma_r(E-z|\vec{E})^{(-1)^{r}}\,,
\ee
where $E=\omega_1+\ldots+\omega_r$.  For simplicity we will denote $S_3( X|\vec{E})=S_3( X)$ and $S_2( X|\vec{Q})=S_2( X)$. 
We also have that:
\be
\label{s2fac}
S_2(X_A)=\Big| \Big| (A;q)\Big| \Big|_{S}^2 e^{\frac{i\pi}{2} B_{22}(X_A)}
\ee
where $A=e^{2\pi i X_A/\omega_2}$ and
\be
B_{2,2}(X)=\frac{1}{6\omega_1\omega_2}\left(6X^2-6(\omega_1+\omega_2)X+\omega_1^2+\omega_2^2+3\omega_1\omega_2\right).\ee

An important property is
\be
S_r(z+\omega_i|\vec{E})=S_{r-1}(z|\vec{E}'_i)^{-1} S_r(z|\vec{E})\,,
\ee
where $\vec{E}'_i=(\omega_1,...\omega_{i-1},\omega_{i+1}, ..\omega_r)$.
 In particular we have:
\be
\label{s3ratio}
\frac{S_3( X+\omega_3|\vec{E})}{S_3(X|\vec{E})}
=S_2(X| \vec{E}')^{-1}=s_b(-i X + i Q/2)^{-1}
\ee
where we set $E'=Q$.  The $s_b(x)$ is the double-sine function. Its  explicit product representation is given by 
\be
s_b(x)=\prod_{m,n\geq 0} \frac{m b +n/b + Q/2-i x}{m b +n/b + Q/2+i x}\, ,
\ee
and satisfies the following identities
\ben
\label{s2c}
s_b(x)s_b(-x)=1\, , \qquad s_b(ib/2-x) s_b(ib/2+ x)=\frac{1}{2 \cosh\pi b x}\, .
\een

\subsection{$\Upsilon^\beta(X)$ function}\label{upsi}
The function $\Upsilon^\beta(X)$ is defined by the following regularized infinite product \cite{Kozcaz:2010af,Bao:2011rc}
\ben\label{qupsi}
\Upsilon^\beta(X)\nn
\propto \prod_{n_1,n_2=0}^{\infty}    \sinh\left[\frac{\beta}{2}\left(X+n_1 b_0+n_2 1/b_0\right)\right]\sinh\left[\frac{\beta}{2}\left(-X+(n_1+1)b_0+(n_2+1)1/b_0\right)\right]\,\\
\een
and satisfies the fundamental properties $\Upsilon^\beta(X)=\Upsilon^\beta(Q_0-X)$ and 
\ben
\label{upp1}
\frac{\Upsilon^\beta(X+b_0)}{\Upsilon^\beta(X)} =[e^{\beta (X-1/(2b_0)) } ]^\infty
 \frac{(e^{\beta /b_0}e^{-\beta  X} ;e^{\beta/b_0)})_\infty }{(e^{\beta X};e^{\beta/b_0})_\infty }= [x q^{-1/2 }]^\infty
 \frac{(q \tilde x ;q)_\infty }{(x;q)_\infty }=
\frac{[xq^{-1/2 }]^\infty}{||(x;q)_\infty ||^2_{\rm id}}\,.\nn\\
\een
where $q=e^{\beta/b_0}$ and $x=e^{\beta X}$, $\tilde x=\bar x=x^{-1}=e^{-\beta X}$.
 
Using the formula $\frac{\sinh \pi x}{\pi x}=\prod_{n=1}^\infty\left(1+\frac{x^2}{n^2}\right)$, the $\Upsilon^{\beta}(X)$ can be related to the standard  $\Upsilon(X)$ as
\be\label{relupsi}
\Upsilon^\beta(X)\propto\prod_{k=-\infty}^{+\infty}\Upsilon\left(X+i\frac{2\pi}{\beta}k\right)\,,
\ee
where 
\be\label{nupsi}
\Upsilon(X)\propto \prod_{n_1,n_2=0}^{\infty}   \left(X+n_1 b_0+n_2 1/b_0\right)\left(-X+(n_1+1)b_0+(n_2+1)1/b_0\right)\,.
\ee

 \section{Basic ($q$-deformed) hypergeometric functions}
The basic (q-deformed) hypergeometric function is represented by the following series
\be\label{qhyper}
\phantom{|}_{n+1}\Phi_n(a_1,\ldots,a_{n+1};b_{1},\ldots,b_{n};z)=\sum_{k=0}^{+\infty} \frac{ (a_1;q)_k\ldots(a_{n+1};q)_k }{(b_1;q)_k\ldots(b_{n};q)_k(q)_k }z^k
\ee
where the $q$-Pochhammer symbols are defined by
\be
(a;q)_k=\prod_{l=0}^{k-1}(1-q^l a),\qquad (q)_k=\prod_{l=1}^{k}(1-q^l).
\ee
In the main text we construct basis of solutions of the $q$-hypergeometric equation using the $\phantom{|}_2\Phi_1(a,b;c;z)$ series
\be\label{qhyper21}
\phantom{|}_2\Phi_1(a,b;c;z)=\sum_{k=0}^{+\infty} \frac{ (a;q)_k (b;q)_k }{(q)_k (c;q)_k}z^k\,.
\ee
The analytic continuation for basic hypergeometric  $\phantom{|}_2\Phi_1$  for $|q|<1$ reads \cite{grahman}
\ben
\label{cont}
\phantom{|}_2\Phi_1(a,b;c;z)&=&\frac{(b;q)_\infty (c/a;q)_\infty }{( c;q)_\infty (b/a;q)_\infty } \frac{(az;q)_\infty (q/(az);q)_\infty }{( z;q)_\infty (q/z;q)_\infty } \phantom{|}_2\Phi_1(a,qa/c;q a/b; c q/(abz)   )+\nn \\ 
&&+
\frac{(a;q)_\infty (c/b;q)_\infty }{( c;q)_\infty (a/b;q)_\infty } \frac{(bz;q)_\infty (q/(bz);q)_\infty }{( z;q)_\infty (q/z;q)_\infty } \phantom{|}_2\Phi_1(b,qb/c;q b/a; c q/(abz) )\, .\nn\\
\een

\bibliography{refs}

\end{document}